\documentclass[reprint,superscriptaddress, amsmath,amssymb,aps, prb, floatfix, compress]{revtex4-2}
\usepackage[defaultcolor=red]{changes}
\setdeletedmarkup{\textcolor{blue}{\sout{#1}}}
\usepackage{graphicx}
\usepackage{dcolumn}
\usepackage{float}
\usepackage{bm}
\usepackage[table]{xcolor}
\usepackage{xcolor}
\usepackage{hyperref}
\usepackage{makecell}
\usepackage{longtable}
\usepackage[T1]{fontenc}
\hypersetup{
    colorlinks,
    linkcolor={red!50!black},
    citecolor={blue!50!black},
    urlcolor={blue!80!black}
}
\begin{document}
\title{A Strategy Toward Room Temperature Topological Hall Effect \\ via Local Moment Magnetism}%

\author{Karthik Rao}
\affiliation{%
Department of Physics and Astronomy, Rice University, Houston, Texas 77005, USA
}%
\affiliation{%
Rice Center for Quantum Materials, Rice University, Houston, Texas 77005, USA
}%
\affiliation{%
Rice Laboratory for Emergent Magnetic Materials and Smalley-Curl Institute, Rice University, Houston, Texas 77005, USA
}%
\author{Kevin Allen}%
\author{Yuxiang Gao}%
\author{Arushi}
\author{Sanu Mishra}%
\affiliation{%
Department of Physics and Astronomy, Rice University, Houston, Texas 77005, USA
}%
\affiliation{%
Rice Center for Quantum Materials, Rice University, Houston, Texas 77005, USA
}%
\author{Birender Singh}%
\author{Kenneth S. Burch}%
\affiliation{%
Department of Physics, Boston College, Chestnut Hill, MA 02467, USA}%
\author{Liangzi Deng}%
\affiliation{%
Department of Physics and Texas Center for Superconductivity at the University of Houston, Houston, Texas 77204, USA
}%
\author{Shanta R. Saha}
\affiliation{%
Maryland Quantum Materials Center, Department of Physics, University of Maryland, College Park, Maryland, 20742, USA
}%
\author{Johnpierre Paglione}%
\affiliation{%
Maryland Quantum Materials Center, Department of Physics, University of Maryland, College Park, Maryland, 20742, USA
}%
\affiliation{%
Canadian Institute for Advanced Research, Toronto, Ontario, M5G 1Z8, Canada
}%
\author{Minseong Lee}%
\author{Vivien Zapf}%
\affiliation{%
National High Magnetic Field Laboratory, Los Alamos National Laboratory, Los Alamos, New Mexico 87545, USA
}%
\author{Emilia Morosan}%
\email[Corresponding author: E. Morosan ]{em11@rice.edu}
\affiliation{%
Department of Physics and Astronomy, Rice University, Houston, Texas 77005, USA
}%
\affiliation{%
Rice Center for Quantum Materials, Rice University, Houston, Texas 77005, USA
}%
\affiliation{%
Rice Laboratory for Emergent Magnetic Materials and Smalley-Curl Institute, Rice University, Houston, Texas 77005, USA
}%
\date{\today}

\begin{abstract}
Topological spin textures in local moment systems hold great promise for technological applications due to their large magnetic moments, strong spin-orbit coupling (SOC) and high tunability. Finding new spin textures that are stable near room temperature is paramount to maximizing their potential for applications. Here, we provide a strategy for realizing topological spin textures at high temperatures by identifying rare earth (\textit{R}) magnets ordering at or near room temperature. We demonstrate the feasibility of this strategy in one of these magnets, hexagonal Gd$_5$Pb$_3$, which orders at $T_C$ = 285 K. The indication for topological spin textures comes from topological Hall effect (THE), which, in Gd$_5$Pb$_3$, occurs  between T = 100 - 200 K, an order of magnitude higher temperature than in other reported \textit{R}-based systems. Our results present an opportunity to explore the role of SOC, anisotropic exchange, geometric frustration, and magnetic interactions in stabilizing topological spin textures,and provide a pathway toward realizing them near room temperature in \textit{R}-based magnets. 
\end{abstract}

\maketitle

\section*{\label{intro}Introduction}
Topological spin textures are of critical importance for both fundamental science and technological applications. From a fundamental standpoint, they are a realization of electronic states governed by topology rather than symmetry breaking alone. From a technological perspective, topological spin textures hold promise for applications in ultra-dense magnetic storage, advanced spintronic devices, and neuromorphic computing \cite{fertSkyrmionsTrack2013, songSkyrmionbasedArtificialSynapses2020}. These spin textures result from non-coplanar magnetic configurations in real space with a non-zero scalar spin chirality given by  $\textbf{S}_i \cdot (\textbf{S}_j \times \textbf{S}_k)$, where $\textbf{S}_i, \textbf{S}_j, \textbf{S}_k$ are nearest-neighbor spins. When itinerant carriers are coupled with such spin textures, they acquire a Berry phase proportional to the spin chirality. As a result, topological spin textures produce an effective magnetic field which, in turn, gives rise to topological Hall effect (THE) \cite{moyaIncommensurateMagneticOrders2022, uedaTopologicalHallEffect2012, tokuraMagneticSkyrmionMaterials2021, moyaRealspaceReciprocalspaceTopology2023, neubauerCorrelationComplexSpin2025a}.

Stabilizing topological spin textures and THE near room temperature is paramount for technological applications. This calls for systems with large magnetic moments, strong magnetic coupling between the local moments and conduction electrons, and strong spin-orbit coupling (SOC). Rare earths $R$ offer the advantage of large magnetic moments and strong SOC, and also high tunability, which makes them particularly promising. For example, magnetic skyrmions in $R$-based compounds show negligible stray fields, making them desirable for skyrmion-based racetrack memories \cite{ExoticRareEarthbased2023}. However, one drawback of $R$ compounds is that spin textures and THE are observed mainly at low temperatures (tens of Kelvin), an energy scale commensurate with the magnetic coupling in \textit{R} intermetallics \cite{hirschbergerSkyrmionPhaseCompeting2019, kurumajiSkyrmionLatticeGiant2019, khanhNanometricSquareSkyrmion2020, yoshimochiMultistepTopologicalTransitions2024, takagiSquareRhombicLattices2022}.

As a way to mitigate this drawback, we propose a strategy to identify $R$-based materials that can host topological spin textures near room temperature (Fig. 1a): \textbf{(Step 1) Strong Coupling}: large moments, strong conduction electrons-local moment coupling and strong SOC are realized in $R$ intermetallics; \textbf{(Step 2) De Gennes Scaling}: in $R$ series of compounds, T$_{ord}$ is proportional with the de Gennes factor  dG $\sim$ $(\text{g}\textsubscript{J} - 1)^2$J(J+1) \cite{deGennesP.1958TPOR}, which is maximum for $R$ = Gd. Therefore, de Gennes scaling predicts that T$_{ord}$ is maximized for $R$ = Gd (Fig. 1b); \textbf{(Step 3) High T$_\textbf{{ord}}$}: while $R$ compounds usually order at tens of Kelvin, we identified several Gd-based compounds with high T$_{ord}$ (250 K $<$ T$_{ord}~<$ 350 K) (Fig. 1c), and expect that THE, if present, will occur at commensurately high temperatures; \textbf{(Step 4) Competing Interactions}: Topological spin textures are stabilized via various competing interactions: Dzyaloshinskii–Moriya interaction (DMI) in non-centrosymmetric systems, \cite{neubauerTopologicalHallEffect2009}  geometric frustration in triangular or hexagonal lattices, or other effects (\textit{e.g.}, higher order coupling between itinerant electrons and local moments in centrosymmetric square net compounds \cite{takagiSquareRhombicLattices2022}); \textbf{(Step 5) Multiple Magnetic Transitions}: here we choose \textit{geometric frustration} to illustrate our strategy. Hexagonal ferromagnet Gd$_5$Pb$_3$ (Fig. 1d-e) had been shown to have multiple magnetic transitions \cite{marcinkovaStrongMagneticCoupling2015}, and is therefore an ideal candidate to validate our strategy. 

\begin{figure*}[t]
    \centering
    \includegraphics[width=1.0\linewidth]{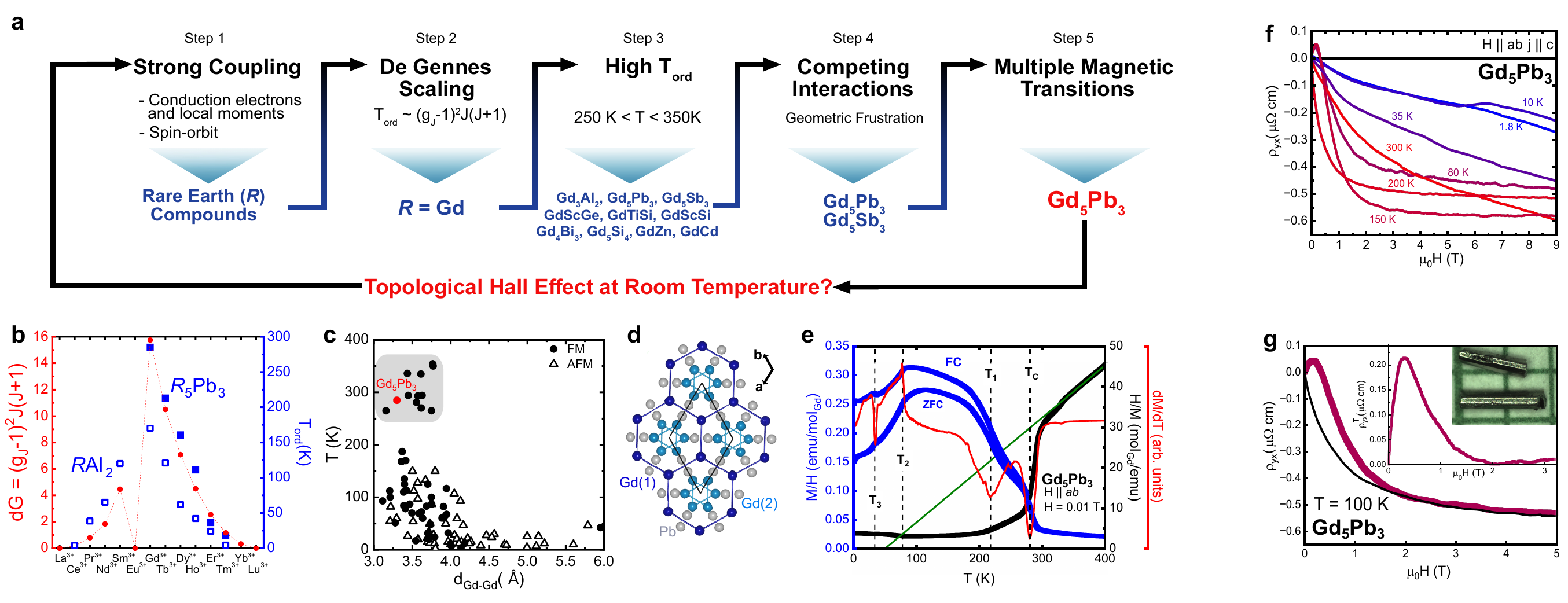}
    \caption{Magnetic phase transitions and topological Hall effect (THE) in Gd$_5$Pb$_3$ (a) Strategy for identifying materials that can have near room temperature topological Hall effect via local moment magnetism (b) De Gennes scaling for rare earth compound series. The ordering temperatures of $R_5$Pb$_3$ compounds are from \cite{marcinkovaStrongMagneticCoupling2015} and the ordering temperatures of $R$Al$_2$ compounds are from \cite{purwinsMagneticPropertiesRare1990}.  (c) $T_\text{ord}$ \textit{vs} shortest Gd-Gd distance in 100 magnetically ordered Gd compounds. The circle (triangle) symbols represent ferromagnetic (antiferromagnetic) transitions, with filled symbols representing members of the Gd$_5M_3$ family. The transition temperatures and distances are listed in Table  S4 in the Supplementary Materials. (d) Top view of the Gd$_5$Pb$_3$ crystal structure consisting of linear chains of Gd(1) atoms (dark blue) along the $c$ axis and a honeycomb array in the $ab$ plane, and face-sharing Gd(2) octahedra (light blue) in the $ab$ plane. The Pb atoms are depicted in gray.   (e) Zero-field cooled (ZFC) and field cooled (FC) DC magnetic susceptibility (left axis, blue), along with their derivative (right axis, red) and inverse magnetic susceptibility (right axis, black) as a function of temperature for H = 0.01 T applied parallel to the $ab$ plane. The green line shows the linear Curie-Weiss fit. (f) Hall resistivity measurements with  $\text{j} || c$ and  $\text{H} || ab$ for temperatures from 2 K to 300 K. (g) Hall resistivity along with the fit (black) using Eq. \ref{hall_rho} at $T = 100 $ K. The inset shows the topological Hall resistivity, and single crystal images with each square equal to $1 \times 1 \text{ mm}^2$.}
    \label{Fig1}
\end{figure*}

 We synthesized single crystals of Gd$_5$Pb$_3$ for the first time and performed a detailed investigation of its electronic, magnetic and structural properties. Gd$_5$Pb$_3$ exhibits a near room temperature ferromagnetic ordering ($T_C=$ 285 K) and undergoes multiple spin reorientation transitions upon cooling in the ordered state. Our magnetization measurements reveal  multiple metamagnetic transitions and a strong uniaxial anisotropy in this Gd$^{3+}$ (L = 0) compound suggesting a non-collinear or non-coplanar magnetic ordering. This is further confirmed by a detailed analysis of Hall resistivity measurements revealing THE throughout most of the ferromagnetic state in Gd$_5$Pb$_3$, with THE maximized around 150 K. These temperatures are the highest among all rare earth systems exhibiting THE \cite{hirschbergerSkyrmionPhaseCompeting2019, kurumajiSkyrmionLatticeGiant2019, khanhNanometricSquareSkyrmion2020, yoshimochiMultistepTopologicalTransitions2024, moyaIncommensurateMagneticOrders2022, moyaRealspaceReciprocalspaceTopology2023}. Our comprehensive investigation validates our proposed strategy in Gd$_5$Pb$_3$ and points to the role of competing exchange interactions between different magnetic sublattices in stabilizing a topologically non-trivial magnetic structure. Our work further establishes that performing similar investigations of the other proposed compounds can provide a pathway for realizing room temperature THE in the local moment limit.

\section*{\label{Methods}Methods}
\subsection*{Sample growth and characterization}
\noindent Single crystals of Gd$_5$Pb$_3$  were grown using a self-flux method from a starting composition of Gd:Pb = 84:16. The mixture was sealed in a tantalum crucible under an argon atmosphere, and sealed in an evacuated quartz tube. The ampoule was heated to 1230 $^\circ$C over 6 hours and kept at this temperature for 2 hours, after which it was cooled to 1180  $^\circ$C over 100 hours. The single crystals were then separated from the excess liquid flux using a centrifuge. Gd$_5$Pb$_3$   forms rod-like crystals with a hexagonal cross-section corresponding to the crystallographic \textit{ab} plane [Fig. 1(g), inset]. The crystals were air-sensitive. Single crystals of the non-magnetic analog, La$_5$Pb$_3$, were grown in a similar manner from a starting composition of La:Pb = 80:20, which was heated to 1200 $^\circ$C and decanted at 800 $^\circ$C.

Powder X-ray diffraction measurements were performed in a  Bruker D8 Diffractometer with Cu K$\alpha$ radiation using an air-tight sample holder. Rietveld structural refinement was performed using FULLPROF software (Supplementary Fig. S1) \cite{rodriguez-carvajalRecentAdvancesMagnetic1993}. The structural parameters for Gd$_5$Pb$_3$  are listed in Supplementary Tables S1 and S2. The crystal orientation was determined using a Laue Camera (Supplementary Fig. S1). The chemical elements of our 
sample were characterized by energy dispersive X-ray spectroscopy in an EMC Helios 660 scanning electron microscope (Supplementary Fig. S2). The elemental analysis indicates the chemical elements of our sample are in good agreement with the chemical formula Gd$_5$Pb$_3$ (Supplementary Table S3). 

\subsection*{Electrical transport measurements}
The ac electrical resistivity measurements were performed using the electrical transport option in a Quantum Design (QD) Physical Property Measurement System (PPMS) DynaCool in a standard four-probe geometry. These measurements were carried out with an applied ac current \textit{j} = 1 mA along the crystallographic $c$ axis for the temperature range of 2 - 300 K, with an applied magnetic field of up to 14 T along [120]. The magnetotransport results were reproduced on a second sample from a different growth batch (Supplementary Fig. S11).

\subsection*{Thermodynamic measurements}
Heat capacity measurements between 1.8 K and 300 K were performed using a thermal-relaxation method in a QD PPMS DynaCool. DC magnetization and magnetic susceptibility measurements between 1.8 K and 400 K, with fields up to 7 T ($H || [001]$ and $H || [120]$) were performed using the Superconducting QUantum Interference Device - Vibrating-Sample Magnetometer (SQUID-VSM) option in a QD magnetic property measurement system (MPMS3). The high-field measurements were performed at the National High Magnetic Field Laboratory at Los Alamos, with fields up to 65 T.

\subsection*{Raman spectroscopy}
Raman scattering measurements were performed in a backscattering geometry using a custom-built optical setup \cite{tianLowVibrationHigh2016}. To avoid any contamination and oxidation, the samples were transferred directly from an argon-filled glovebox to the low-temperature Raman cryostat using a high-vacuum (1.0 $\times 10^{-8}$  Torr) transfer suitcase \cite{grayCleanroomGlovebox2020}. A 532 nm (2.33 eV) laser was focused onto the sample surface using a 100$\times$ long-working-distance objective. The laser power was kept below 0.2 mW to minimize local heating. The Raman scattered signal was dispersed using a 1200 grooves per millimeter grating and detected with an Andor spectrometer equipped with a charge-coupled device (CCD) detector. The sample temperature was controlled between 10 K and 320 K using a closed-cycle continuous He-flow cryostat (Montana Instruments) under high vacuum (1.0 $\times 10^{-8}$  Torr). The different symmetry channels were probed via controlling the incident and scattered light polarization using half- and quarter-wave plates in combination with a fixed analyzer.

\section*{\label{results}Results and Discussion}

\subsection*{Room Temperature Ferromagnetism}
 \begin{figure*}
    \centering
    \includegraphics[width=1.0\linewidth]{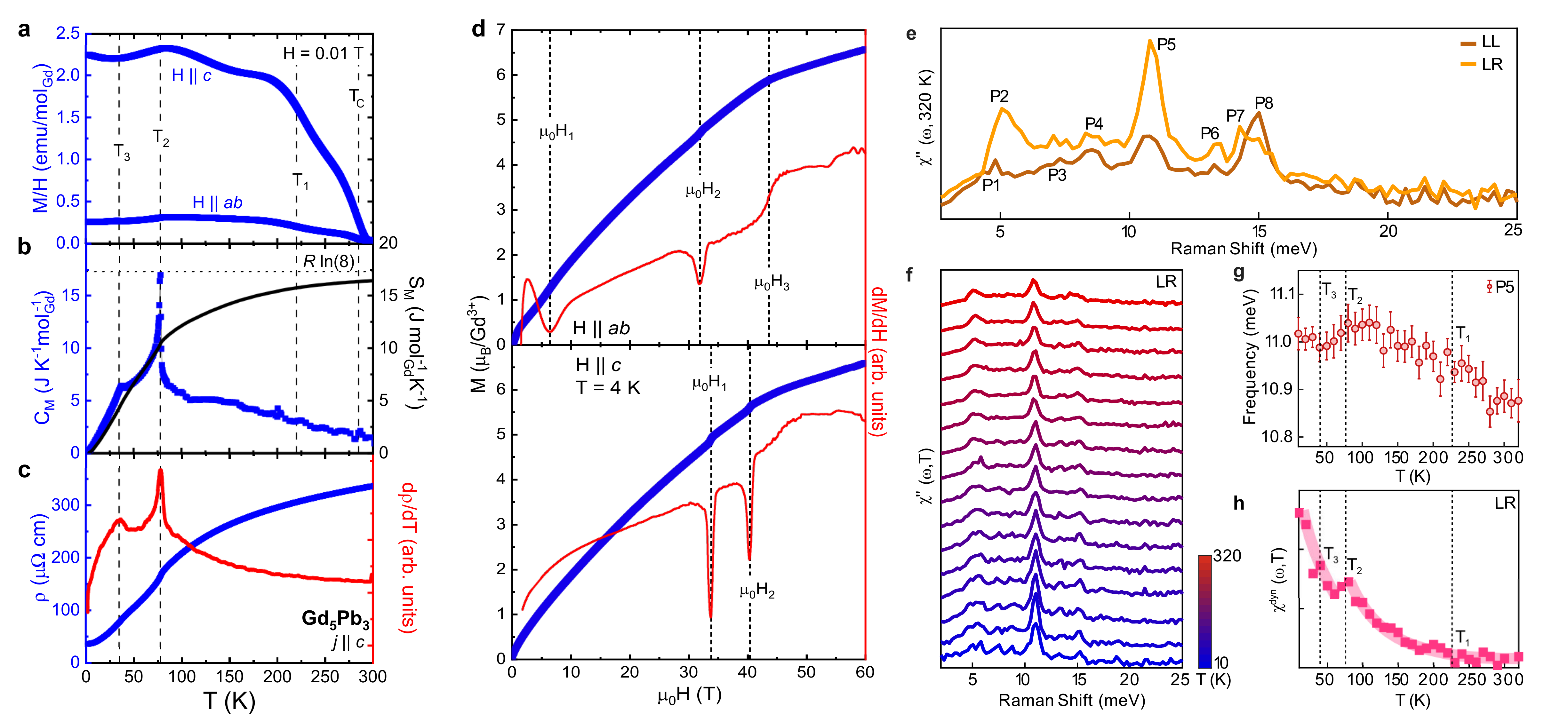}
    \caption{(a) DC magnetic susceptibility for a 0.01 T field applied parallel to the $c$ axis and parallel to the $ab$ plane. (b) Temperature-dependent magnetic contribution to the specific heat (black, left axis) and the magnetic entropy (red, right axis). (c) Electrical resistivity as a function of temperature (black, left axis) along with its derivative (red, right axis) for $j || c$. The dashed lines represent the phase boundaries. (d) Magnetization measurements performed up to 60 T (blue, left axis) along with the derivative (red, right axis) for field parallel to the $ab$ plane (top) and parallel to the $c$ axis (bottom). (e-h), Raman susceptibility $\chi''$ ($\omega$,$T$) of Gd$_5$Pb$_3$ measured at 320 K in co (LL) and crossed (LR) circular polarization channel (e), Temperature evolution of $\chi''$ ($\omega$,$T$) in the LR symmetry channel from 10 K to 320 K, with the color bar indicating temperature (f) Temperature dependence of the energy of the phonon mode at 11 meV (P5) (g), and the dynamic Raman susceptibility $\chi^{\text{dyn}}$($\omega$,$T$) (h), the vertical dashed lines indicate the magnetic transition temperatures, and the pink solid line in (h) is a guide to the eye. The error bars were obtained from the least-squares fitting of the Raman spectra and represent the statistical uncertainties associated with the fitted mode frequency.
}
    \label{Fig2}
\end{figure*}

A previous study on hexagonal $R_5$Pb$_3$ compounds \cite{marcinkovaStrongMagneticCoupling2015} confirms the de Gennes scaling across the series with $R$ = Gd - Tm (steps 1-2). In the current study, DC magnetic susceptibility \textit{M(T)/H} was measured on single crystals of Gd$_5$Pb$_3$, with an external field H = 0.01 T applied along the $ab$ plane (Fig. 1e, left axis) and the $c$ axis (Supplementary Fig. S5c). Upon cooling, the susceptibility increases sharply below $T_C \approx 285$ K, indicating ferromagnetic ordering (step 3), followed by another sharp increase at $T_1 \approx 220$ K.  On further cooling, two more transitions occur at $T_2 \approx 78$ K and $T_3 \approx 34$ K, marked by cusps in the M/H (blue) and dM/dH (red) curves in Fig. 1e (step 5). The decrease in the magnetic susceptibility with decreasing temperature is indicative of spin reorientation phase transitions. Gd$_5$Pb$_3$ is one of several $R_5M_3$ compounds ($R$ = rare earth or Mn, $M$ = Si, Ge, Sn, Sb, Bi, Pb) showing multiple magnetic transitions and complex H - T phase diagrams \cite{marcinkovaStrongMagneticCoupling2015, szadeMagnetismElectronicStructure2000, morozkinMagneticStructureCompound2008, kitaoriMagneticPropertiesSingle2023, mondalCompetingMagneticInteractions2022, samathamMagnetizationResistivitySpecific2018, nagaiMagneticElectricalProperties2007, schobingerpapamantellosMagneticOrderingIncommensurate1982, semitelouMagneticStructureTb5Sn32000, kitaoriEnhancedEmergentElectromagnetic2024, semitelouMagneticStructureTb5Sn32000, semitelouSinemodulatedMagneticStructure1992, semitelouCommensurateIncommensuratePhases2003, semitelouAmplitudemodulatedSpinStructure2000, kurisuDenseKondoCompound1999, lawrenceCoexistenceMagneticOrder1991, yanMagneticOrderNd52018}. These $R_5M_3$ compounds crystallize in the centrosymmetric hexagonal Mn$_5$Si$_3$-type structure (space group $P6_3/mcm$) with two inequivalent $R$ crystallographic sites \cite{marcinkovaStrongMagneticCoupling2015}: the $R$(1) atoms (located at site 4\textit{d}; Fig. 1d, dark blue)  form linear chains along the \textit{c} axis, with the chains ordered in a honeycomb array; $R$(2) atoms (located at site 6\textit{g}; Fig. 1d, light blue) form face-sharing trigonal antiprisms along the \textit{c} axis, arranged in a hexagonal array. Therefore, the physical properties of Gd$_5$Pb$_3$ are dictated by the coupling between the two different magnetic sublattices, leading to anisotropic exchange and possible geometric frustration. This explains the complex magnetic behavior with multiple magnetic and spin reorientation transitions between non-trivial magnetic configurations (step 5).

The zero-field cooled (ZFC) and field cooled (FC) magnetic susceptibility curves split at temperatures below $T_C$, indicating hysteresis commonly associated with ferromagnetic transitions. After subtracting a temperature-independent term $M_0/H$, a Curie-Weiss \cite{Mugiraneza_Hallas_2022} linear fit (Fig. 1b) on the inverse susceptibility data between 330 K and 400 K yields a Weiss temperature $\theta_W = 52.5$ K, indicating ferromagnetic coupling, and an effective paramagnetic moment $\mu^{\text{exp}}_\text{eff} \approx 7.87 \, \mu_\text{B}$/Gd$^{3+}$ ion, close to the theoretical effective magnetic moment $\mu^{\text{theory}}_\text{eff} \approx 7.94 \, \mu_\text{B}$. The ratio $\theta_W/T_C \approx 0.2 \ll 1 $ suggests a rare ``anti-frustration" effect where the $T_C$ is several times larger than $\theta_W$. This type of effect has been observed in other members of the $R_5\text{Pb}_3$ family \cite{marcinkovaStrongMagneticCoupling2015}, and a few other rare earth compounds such as GdPtPb \cite{manniGdPtPbNoncollinearAntiferromagnet2017} and EuRh$_2$As$_2$ \cite{singhUnusualMagneticThermal2009} due to anisotropic exchange couplings both within a magnetic sublattice and between different magnetic sublattices. These competing exchange interactions within and between the two Gd magnetic sublattices are likely responsible for the multiple magnetic transitions in Gd$_5$Pb$_3$. Given that Gd$_5$Pb$_3$ meets all criteria outlined in steps 1-5, we now explore in detail its magnetotransport properties in search for signatures of THE. Fig. 1f shows Hall measurements from 1.8 K (T $<$ T$_{ord}$) to 300 K (T $>$ T$_{ord}$). High temperature, non-zero topological Hall resistivity is evident in Gd$_5$Pb$_3$, as illustrated in Fig. 1g for T  = 100 K and discussed in detail below.

The $T_C$ is remarkably high for a rare earth compound, and implies strong exchange coupling exists in Gd$_5$Pb$_3$. This makes Gd$_5$Pb$_3$ one of the very few compounds with rare earth-only magnetic moments that order close to room temperature (Fig. 1c). Supplementary table 4 lists the magnetic ordering temperatures along with the shortest Gd-Gd distance for 100 compounds where Gd is the only magnetic ion. The vast majority of known Gd compounds order at temperatures below 100 K; however, there is a cluster of 10 ferromagnetic compounds (and elemental Gd) which have ordering temperatures above 250 K (Fig. 1c, grey area). The low ordering temperatures are expected due to the localized nature of the $4f$ electrons and their relatively weak exchange interactions. Several of the compounds in the cluster near Gd$_5$Pb$_3$ have been reported to have multiple magnetic transitions \cite{samathamMagnetizationResistivitySpecific2018, zhaoLargeAnomalousHall2021, pecharskyGiantMagnetocaloricEffect1997, szadeMagnetismElectronicStructure2000, skorekElectronicStructureMagnetism2001} and yet extensive magnetotransport studies on single crystals are lacking. Therefore, detailed investigations of these high temperature magnetically ordered Gd compounds are currently underway.

\subsection*{Magnetic Anisotropy in \texorpdfstring{$\mathbf{Gd_5}\mathbf{Pb_3}$}{Gd5Pb3}}

Our magnetization measurements reveal a strong magnetic anisotropy, which is unusual for a Gd-based compound where crystalline field anisotropy is expected to be negligible. The DC magnetic susceptibility measured with the field applied along the $c$ axis is significantly greater below $T_c$ than that for $H| | ab$ (Fig. 2a). Additionally, the susceptibility continues to increase below the transition at $T_3$ for $H || c$ and decreases below $T_3$ for $H|| ab$. Upon increasing the applied field to 7 T, the transition $T_1$ decreases slightly for both directions, and $T_3$ increases slightly only for the field applied along the $c$ axis, while $T_C$ and $T_2$ remain unaffected (Supplementary Fig. S5a-j). The transitions are not suppressed with increasing field up to 7 T. 

This anisotropy is also evident in the magnetization isotherms $M(H)$ measured at high magnetic fields (Fig. 3d). The magnetization isotherms measured for Gd$_5$Pb$_3$ at various temperatures between 2 K and 300 K increase linearly with the applied magnetic field between 0.2 T and 7 T (Supplementary Fig. S3a, b), and the maximum magnetization reached in this field range is  $\approx 1.5 \, \mu_\text{B}$ per Gd$^{3+}$, which is much smaller than the expected saturation moment for Gd$^{3+}$ ions ($\mu_\text{sat} = 7.0 \mu_\text{B}/\text{Gd}^{3+}$). Small hysteresis is observed in the magnetization curve measured at low temperatures with $H || ab$ for $|\mu_0H| < 0.6$ T (Supplementary Fig. S3c, d), which indicates Gd$_5\text{Pb}_3$ is a soft ferromagnet. The magnetization approaches the expected saturation moment of Gd$^{3+}$ at high magnetic fields ($\mu_0H \approx$ 60 T). There are up to three meta-magnetic transitions observed in the magnetization isotherms below $T_2$ (Supplementary Fig. S4a-j) and the critical fields for the metamagnetic transitions are different: for $H || ab$, the critical fields are $\mu_0H_1 \approx 6.4 $ T, $\mu_0H_2 \approx 31.8 $ T, $\mu_0H_3 \approx 44.3 $ T  at 4 K, and $\mu_0H_1 \approx 8.2 $ T at 75 K; for $H || c$, the critical fields are $\mu_0H_1 \approx 33.7 $ T and $\mu_0H_2 \approx 40.4 $ T at 4 K, and there is no metamagnetic transition at 75 K below the maximum measured field of 60 T (Supplementary Fig. S4). Anisotropy in the magnetically ordered state has been observed in other Gd compounds, including the skyrmion lattice compounds, and is usually indicative of a complex magnetic ordering such as a helical, amplitude-modulated, or other non-trivial magnetic structure \cite{garnierGiantMagneticAnisotropy1996, khanhNanometricSquareSkyrmion2020, nakamuraMagneticPhasesFrustrated2023, sahaMagneticAnisotropyFirstorderlike1999}. For instance, the effective four-spin interaction led to magnetic anisotropy and stabilized a helical magnetic order and the skyrmion lattice in GdRu$_2$Si$_2$ \cite{khanhNanometricSquareSkyrmion2020}. Anisotropy in Gd$_3$Ru$_4$Al$_{12}$ was caused by the magnetic quadrupoles of the ferromagnetic trimers \cite{nakamuraMagneticPhasesFrustrated2023}. In elemental Gd, the magnetocrystalline anisotropy was attributed to the spin-orbit interaction from the 5$d$-6$s$ conduction electrons resulting in a valence band moment due to the splitting induced by the exchange field from the 4$f$ local moment. This spin-orbit coupling, and the subsequent anisotropy was dependent on the magnitude of the 4$f$ moment \cite{colarieti-tostiOriginMagneticAnisotropy2003}. As such, the magnetic anisotropy we observe could result in a helical or other non-coplanar magnetic configurations in Gd$_5$Pb$_3$.
\begin{figure*}
    \centering
    \includegraphics[width=1.0\linewidth]{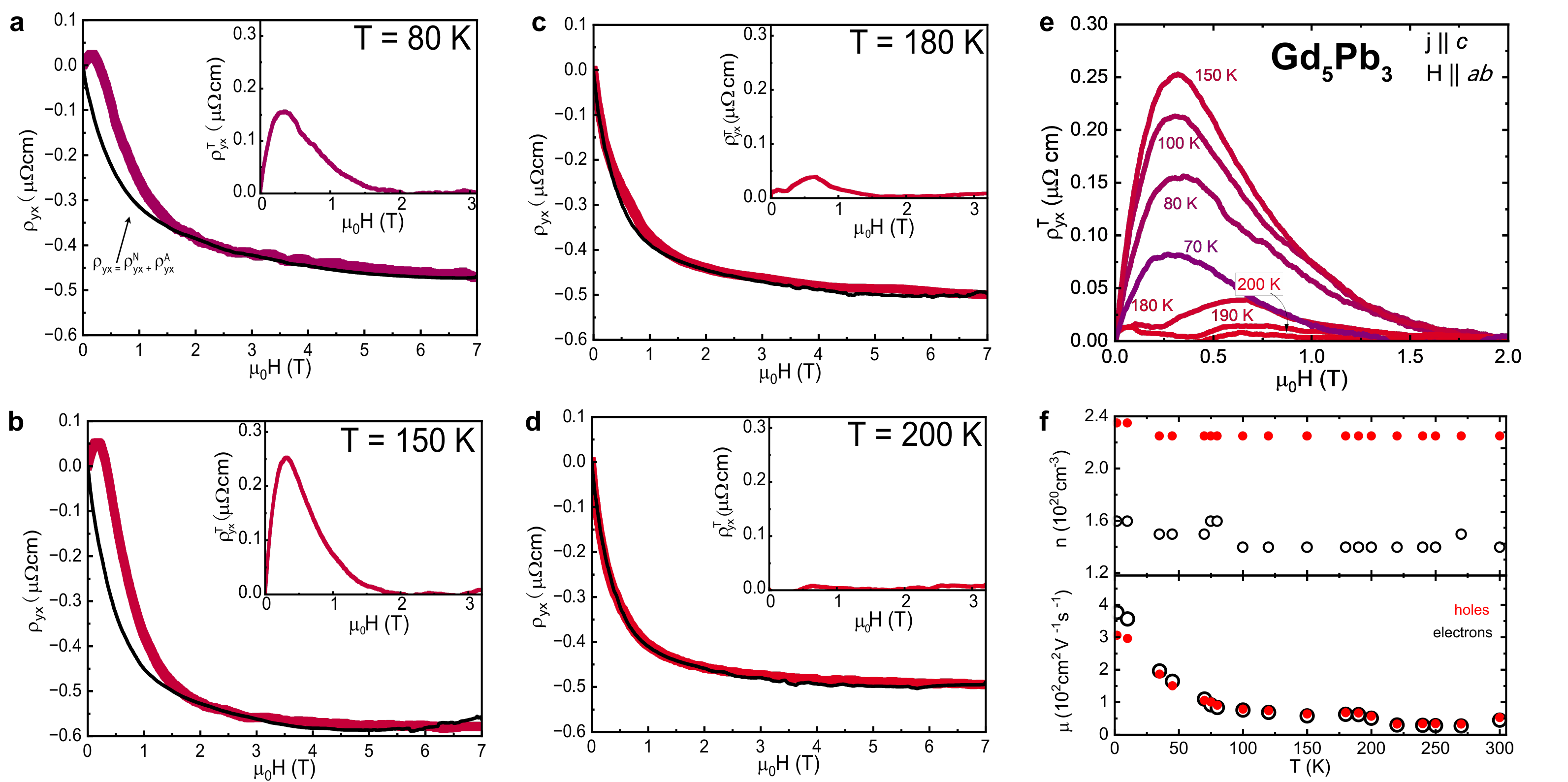}
    \caption{ Hall resistivity along with the fit (black) using Eq. \ref{hall_rho} at $T = 80 $ K (a), 150 K (b), 180 K (c), and 200 K (d). The inset shows the topological Hall resistivity, obtained by subtracting the measured curve and fit. (e) The topological Hall resistivity in Gd$_5$Pb$_3$. (f) Temperature-dependent carrier concentrations (top) and  mobilities (bottom) for holes (red filled circles) and electrons (black open circles).}
    \label{Fig3}
\end{figure*}

\subsection*{Magnetic Phase Transitions in \texorpdfstring{$\mathbf{Gd_5}\mathbf{Pb_3}$}{Gd5Pb3}}
In the heat capacity of Gd$_5$Pb$_3$ and its non-magnetic analog, La$_5$Pb$_3$, (Supplementary Fig. S6) the phonon contribution can be described by a Debye-Einstein model (for further details, see Supplementary Material). In the magnetic specific heat $C_m$, obtained by subtracting the scaled specific heat of La$_5$Pb$_3$  from that of Gd$_5$Pb$_3$ \cite{Bouvier_Lethuillier_Schmitt_1991}, two sharp peaks indicate the transitions at $T_2 = 78 $ K and $T_3 = 34 $ K, in agreement with the magnetic susceptibility measurements  (Fig. 2b, left axis). Specific heat measurements are difficult to perform above 100 K. The magnetic entropy approaches the expected \textit{R}ln(8) close to $T_C$, indicating an eight-fold degenerate ground state as  expected for the Gd$^{3+}$ ion ($J = 7/2$) (Fig. 2b, right axis).

To determine whether the phase transitions observed have a structural component, we performed temperature-dependent Raman measurements across a wide temperature range. The group theory analysis for the $P6_3/mcm$ (No. 193) space group predicts a total of ten,  $\Gamma_{\text{Raman}} = 2 \text{A}_{1g}+5 \text{E}_{2g}+3\text{E}_{1g}$, phonon modes in the irreducible representation, which necessitates probing the different symmetry channels. Fig. 2e displays the Raman susceptibility, $\chi''$ ($\omega$,$T$), measured at 320 K under co-circular (LL) and crossed-circular (LR) polarization, where $\omega$ is the phonon energy and L(R) denotes left- (right-) circularly polarized light. Eight phonon modes labeled P1-P8 (see Fig. 2e) are clearly observed. Fig. 2f illustrates the full temperature evolution of $\chi''$ ($\omega$,$T$) in the LR channel, revealing no noticeable change in the spectra and no additional phonon modes between 10 and 320 K, thus suggesting the absence of any structural phase transition. Similarly, no additional phonon modes are observed in spectra in the LL channel between 10 to 320 K (Supplementary Fig. 7a). 

Furthermore, the measured Raman responses provide evidence for significant magneto-elastic coupling. We tracked the temperature dependence of the energy for the prominent phonon at 11 meV (P5), as shown in Fig. 2g, where the energy is extracted from the Lorentzian fits. The phonon energy increases gradually upon cooling, consistent with standard anharmonic shifts. However, it exhibits a small softening near $T_2$, and remains nearly constant below $T_3$, while no anomalous change in the mode energy is observed across $T_1$. An analogous softening is observed for the 15.2 meV (P8) phonon mode (Supplementary Fig. 7b), indicating a coupling of the lattice to the magnetic order emerging at $T_2$. 

We also used Raman measurements to explore the magnetic fluctuations in Gd$_5$Pb$_3$. In particular, the broad feature extending to 20 meV could originate from electronic or magnetic fluctuations. However, for the free carrier response, no substantial temperature dependence is expected. If this resulted from magnetic fluctuations, it should be directly tied to the magnetic transitions. With this in mind, we evaluated the dynamic Raman susceptibility $\chi^{\text{dyn}}$($\omega$,$T$), which probes the integrated spectral weight of collective excitations. The dynamic susceptibility $\chi^{\text{dyn}}$($T$) is obtained by integrating the Raman conductivity $\chi''$ ($\omega$)/$\omega$ (see Supplementary Fig. 7d and Fig. 7e) up to an upper cutoff of 25 meV, above which the spectral background shows no temperature dependence, using the Kramers-Kronig relation \cite{gallaisChargeNematicityElectronic2015},  $$ \chi^{\text{dyn}}(\omega,T) = \frac{2}{\pi}\int_{0}^{25 \text{ meV}} \frac{\chi'' (\omega)}{\omega} d\omega. $$ The temperature response of the dynamic susceptibility $\chi^{\text{dyn}}$($T$), as shown in Fig. 2h, remains nearly constant above $T_1$, and increases gradually between $T_1$ and $T_2$. Interestingly, a change in slope is observed at $T_2$, but further work is needed to determine if this results from changes in the phonons or magnons at this temperature. However, we found an identical trend in the dynamic susceptibility $\chi^{\text{dyn}}$($T$)  obtained from the LL symmetry channel (Supplementary Fig. 7c). The rapid increase in $\chi^{\text{dyn}}$($T$) below $T_1$ and not below $T_C$ is surprising but could be consistent with $T_C$ being associated with ferromagnetic order, while $T_1$ results from the onset of a more complex magnetic structure. These results are also consistent with the lower temperature magnetic transitions being subtle spin reorientation that does not dramatically change the fluctuation spectra.

Noting the strong magnetic fluctuations detected by Raman, we turned to electrical transport, which is likely to be affected by the nontrivial symmetry of the magnetic order. Due to the crystal geometry, size and air-sensitivity, we were only able to perform magnetotransport measurements with current along the $c$ axis, and field perpendicular to the $c$ axis. Furthermore, we were not able to measure resistivity at sufficiently high temperatures above $T_C$ as the sample surface started to degrade above 320 K. The zero-field electrical resistivity $\rho(T)$ of Gd$_5$Pb$_3$  shows metallic behavior, with $\rho$ decreasing with cooling from 300 K to 2 K (Fig. 2c). The residual resistivity ratio (RRR), given by $\text{RRR} = \rho_\text{300 K}/\rho_\text{2 K}$, is 5. The $T_2 = 78 $ K and $T_3 = 34 $ K transitions are confirmed by the resistivity measurements, with corresponding peaks in $d\rho/dT$ (red curve, Fig. \ref{Fig2}c). As the Raman scattering shows no structural transitions, and the Hall coefficient (discussed later) has no abrupt changes across these transitions, these features in resistivity are likely associated with the magnetic phase transitions. An explanation for the lack of change in the resistivity slope at $T_C$ and $T_1$ is that the charge carriers are not affected by the spin ordering or reorientation because the current carrying direction is along the easy axis of the material. The resistivity has an atypical behavior above $T_2$, where it increases slower than the expected linear scaling ($\rho \propto T$) due to phonon scattering at high temperatures. This behavior is also seen in the non-magnetic analog, La$_5$Pb$_3$ (Fig. S8) down to lower temperatures, where a model that took into account scattering from acoustic and optical phonons was used to fit the resistivity (further details in the Supplementary Material).

\subsection*{Topological Hall Effect}
The Hall resistivity in magnetic materials hosting topological spin textures can be expressed as the sum of normal Hall resistivity ($\rho^N_{yx}$), anomalous Hall resistivity ($\rho^{A}_{yx}$) and topological Hall resistivity ($ \rho^{T}_{yx}$) \cite{moyaIncommensurateMagneticOrders2022} as described in Eq. \ref{hall_rho}:
\begin{equation}\label{hall_rho}
    \rho_{yx} = \rho^N_{yx} + \rho^{A}_{yx} + \rho^{T}_{yx} = R_0\mu_0H + S_H\rho^n_{xx}M +  \rho^{T}_{yx}, 
\end{equation}
where $R_0$ is the normal Hall coefficient for a one-band or two-band model, $S_H$ is the anomalous Hall coefficient, $M$ is the magnetization, $\rho_{xx}$ is the magnetoresistance, with $n = 2 $ or $ 1 $ depending on a dominant intrinsic mechanism or a skew scattering mechanism, respectively \cite{nagaosaAnomalousHallEffect2010}. In the high-field region where the magnetization saturates, topological spin structures are expected to become trivial spin structures, and the topological Hall resistivity vanishes \cite{kurumajiMetamagneticMultibandHall2024, surgersLargeTopologicalHall2014, moyaRealspaceReciprocalspaceTopology2023}. The topological Hall resistivity can then be determined by subtracting the normal and anomalous Hall resistivity contributions from the measured Hall resistivity.

\begin{figure}[H]
    \centering
    \includegraphics[width=1.0\linewidth]{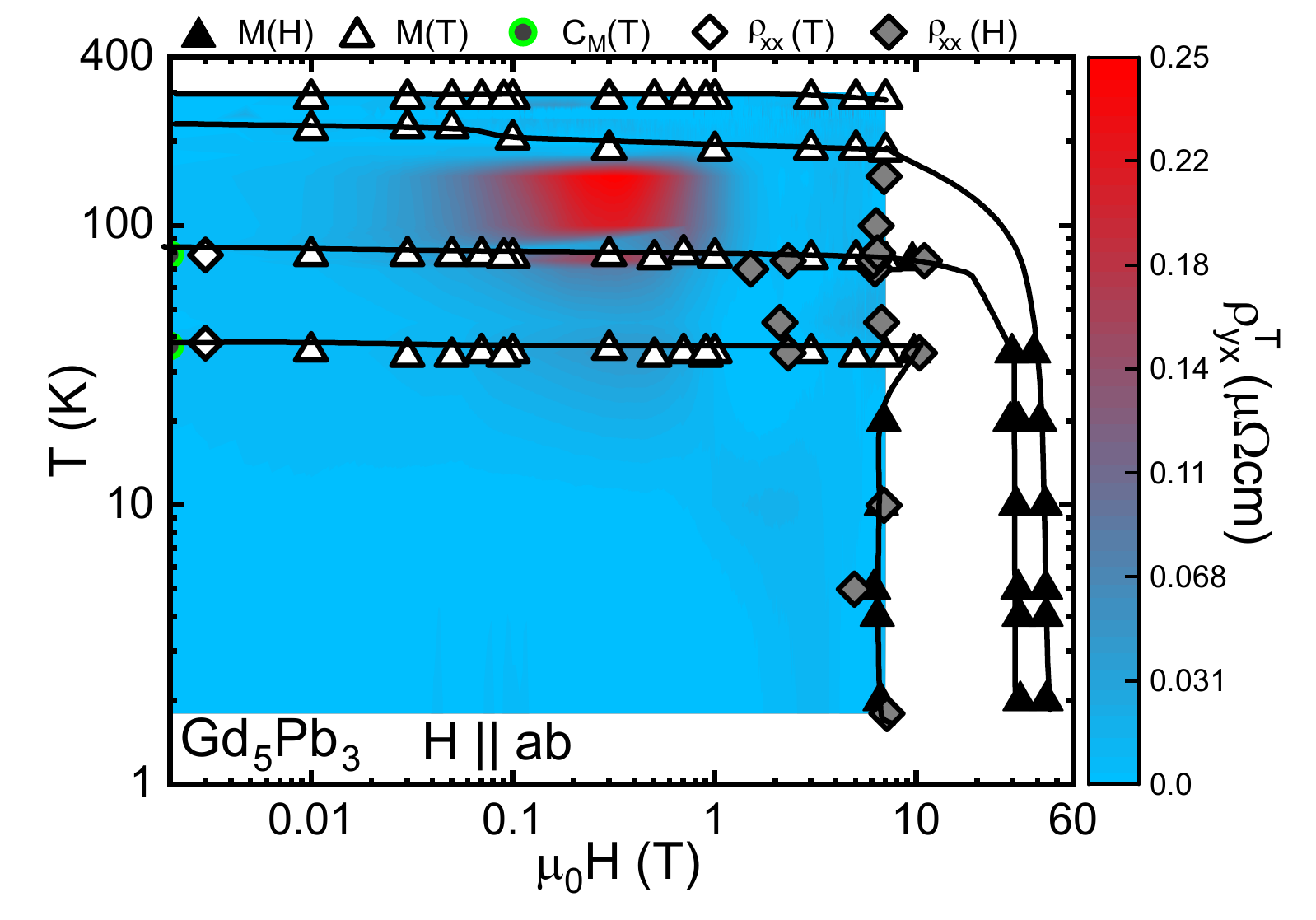}
    \caption{Magnetic $H - T$ phase diagram constructed from peaks in the derivatives of magnetic susceptibility (open triangles), magnetization (filled triangles), magnetic specific heat (green filled circle), resistivity as a function of temperature (open diamonds), resistivity as a function of field (filled diamonds) for $\text{H} || ab$ on a log-log scale. The magnetic specific heat and temperature-dependent resistivity measurements were performed at zero field and have been offset by a small amount to display properly on the log-scale. The contour map is generated using $\rho^{T}_{yx}$.}
    \label{Fig4}
\end{figure}

The field-dependent Hall resistivity $\rho_{yx}(H)$ in Gd$_5$Pb$_3$ has distinct non-monotonic behavior across different transitions (Fig. 1f). At 300 K, in the paramagnetic state, the Hall resistivity is non-linear and cannot be described by a simple one-band model for the normal Hall effect.  For $T_1 < T< T_C$, the Hall resistivity drops sharply to $H =  1$ T and saturates around -0.6 $\mu \Omega$ cm. Between $T_1$ and $T_2$, the Hall resistivity has a strong non-linear dependence on the magnetic field, as it changes sign and peaks slightly less than 0.1 $\mu \Omega$ cm before dropping sharply up to 2 T saturating around -0.5 $\mu \Omega$ cm. Below $T_2$, the Hall resistivity no longer switches sign, and it decreases non-linearly without saturating, before developing a peak around 7 T below $T_3$, which corresponds to the critical field for the peak in magnetoresistance (Supplementary Fig. S9) and the metamagnetic transition in magnetization. 

The magnetization of Gd$_5$Pb$_3$ does not saturate in the magnetic field range where magnetotransport measurements were performed, which makes it non-trivial to extract the various contributions to the Hall resistivity. As the Hall resistivity is non-linear in the paramagnetic state, we used a combination of a two-band model along with a skew-scattering dominant anomalous Hall effect to fit the data and obtain our initial fitting parameters (further details in the supplementary material). We were able to achieve good fits with the Hall resistivity data at all temperatures except between $T_1$ and $T_2$ (Fig. 3a-e), where an enhancement in the measured Hall resistivity cannot be explained using just the intrinsic or extrinsic anomalous Hall effect. The extra contribution is therefore attributed to the THE. 

The carrier concentrations extracted from the two-band model show that the normal Hall contribution is slightly dominated by holes. The concentration remains relatively unchanged over the measured temperature range (Fig. 4f), with the hole concentration ($n_\text{h}$) approximately 2.25 $\times 10^{20}$ cm$^{-3}$ and the electron concentration ($n_\text{e}$) approximately 1.4 $\times 10^{20}$ cm$^{-3}$. The carrier mobilities, on the other hand, increase non-linearly with decreasing temperature, with changes occurring at the magnetic transition. At 300 K, the mobilities ($\mu_{\text{h}}$ for holes, $\mu_{\text{e}}$ for electrons) are $\mu_{\text{h}} \approx 54 \text{ cm}^2 \text{ V}^{-1} \text{ s}^{-1}$ and $\mu_{\text{e}} \approx 46 \text{ cm}^2 \text{ V}^{-1} \text{ s}^{-1}$. Both $\mu_{\text{h}}$ and $\mu_{\text{e}}$ decrease slightly below $T_C$, before increasing again below $T_1$, and further cooling results in a greater increase in both mobilities. The anomalous Hall coefficient $S_H$ (Supplementary Fig. S10k) is nearly unchanged across the transitions and remains between $-7 \times 10^{-8}$ m/A and $-1 \times 10^{-7}$ m/A. Similar behavior of the Hall coefficients had been observed in other systems exhibiting THE such as the isostructural compound Mn$_5$Si$_3$ \cite{surgersLargeTopologicalHall2014}
 and GdRu$_2$Si$_2$ \cite{guptaSkyrmionPhasePolycrystalline2025}. This indicates the presence of a non-trivial spin structure as the chirality-induced internal field acts to cancel the applied field, leading to opposite signs for $R_0$ and $S_H$. 

The longitudinal conductivity can be expressed as $\sigma_{xx} = \rho_{xx}/\left( \rho^2_{xx} + \rho^2_{yx}\right)$. Similarly, the anomalous Hall conductivity can be calculated as $\sigma^{A}_{xy} = \rho^A_{yx}/\left( \rho^2_{xx} + \rho^2_{yx}\right)$. Noting that $\rho_{yx} << \rho_{xx}$ in Gd$_5$Pb$_3$ (Fig. 1d, Supplementary Fig. S9), we obtain $\sigma_{xx}$ between $10^5$ $ \Omega^{-1} \text{cm}^{-1}$ and  $10^6$ $ \Omega^{-1} \text{cm}^{-1}$ and  $\sigma^A_{yx}$ between $10^3$ $ \Omega^{-1} \text{cm}^{-1}$ and  $10^5$ $ \Omega^{-1} \text{cm}^{-1}$. Empirically, for highly conductive systems with $\sigma_{xx} > 10^5$ $ \Omega^{-1} \text{cm}^{-1}$, $\sigma^A_{xy}$ is expected to be dominated by skew scattering and scale linearly with $\sigma_{xx}$ \cite{moyaRealspaceReciprocalspaceTopology2023,nagaosaAnomalousHallEffect2010}, which is observed in Gd$_5$Pb$_3$ (Supplementary Fig. S10l). The tangent of the anomalous Hall angle $\tan(\Theta^A_H) = \left(\sigma^A_{xy}/\sigma_{xx} \right)$ is approximately 0.014, as expected for a typical metal with a dominant skew scattering contribution \cite{moyaRealspaceReciprocalspaceTopology2023, fujishiroGiantAnomalousHall2021}.

The topological Hall resistivity obtained after subtracting the normal and anomalous contributions for temperatures between $T_1$ and $T_2$ is positive, with a peak around 0.3 T, and becomes negligible above 2 T. The maximum $\rho^T_{yx}$ is 0.25 $\mu \Omega \text{ cm}$ obtained at 150 K around 0.3 T, suggesting that the non-coplanar magnetic structure contributing to the THE gradually weakens upon reaching the phase boundaries at $T_1$ and $T_2$. The magnitude of $\rho^T_{yx}$ is comparable to that of other compounds with THE such as Eu(Ga$_{1-x}$Al$_x$)$_4$ \cite{moyaRealspaceReciprocalspaceTopology2023}, Mn$_5$Si$_3$ \cite{surgersLargeTopologicalHall2014} and GdRu$_2$Ge$_2$ \cite{yoshimochiMultistepTopologicalTransitions2024}. However, Gd$_5$Pb$_3$ stands out as this THE occurs at significantly higher temperatures, especially compared to the other rare earth-based compounds, which, combined with room-temperature ferromagnetism, increases the potential for applications in advanced technologies. In Gd-based compounds (L = 0), the presence of multiple magnetic phase transitions is often an indication of non-collinear or non-coplanar magnetic configuration. Furthermore, THE can occur from the  static scalar spin chirality attributed to non-coplanar spin textures, or from the dynamically generated scalar spin chirality due to chiral fluctuations, as seen in the non-coplanar transverse conical spiral phase of YMn$_6$Sn$_6$ \cite{ghimireCompetingMagneticPhases2020} and ErMn$_6$Sn$_6$ \cite{fruhlingTopologicalHallEffect2024}. Therefore, our observation of THE and multiple magnetic phase transitions in  Gd$_5$Pb$_3$ strongly suggests a non-coplanar arrangement of moments between $T_1$ and $T_2$. 

We have mapped out the $H - T$ phase diagram of Gd$_5$Pb$_3$ using the results from our thermodynamic and magnetotransport measurements (Fig. 4 and Supplementary Fig. S12). The points in the phase diagrams are taken from peaks in the magnetic specific heat, the derivatives of magnetic susceptibility and resistivity as a function of temperature, and the derivatives of magnetization and magneto-resistance as a function of magnetic field. The phase diagram reveals several boundaries that separate phases with different magnetic spin configurations. As noted previously, Gd$_5$Pb$_3$ has anisotropic magnetic properties, and this could stem from a helical or other non-coplanar magnetic configuration. The contour map is constructed based on the observed topological Hall resistivity from Fig. 1f, and indicates that the THE contribution is maximum between $T_1$ and $T_2$.

In summary, we present a five-step strategy to identify topological spin textures near room temperature in local moment systems:  in order to enhance the SOC and conduction electron-local moment coupling, as well as the magnetic moments, we focus on $R$ compounds (step 1). Of these, de Gennes scaling ensures that Gd compounds have the highest ordering temperature (step 2). While $R$ compounds usually order at tens of Kelvin, we identify 10 Gd-based intermetallics that order between 250 K and 350 K (step 3), with several of them displaying complex magnetic order as evidenced by multiple magnetic transitions (step 5). The multiple magnetic transitions in Gd compounds (with L = 0) are due to different interactions (frustration, DMI, higher order coupling) competing with the magnetic coupling (step 4). We chose hexagonal Gd$_5$Pb$_3$ as a candidate material to validate our strategy: for the first time, we synthesized single crystals of Gd$_5$Pb$_3$ and performed extensive thermodynamic and electrical transport measurements. We reveal up to four magnetic transitions, with a remarkably large ordering temperature close to room temperature. Furthermore, the detailed $H||ab$ magnetotransport data analysis points to THE contribution in Gd$_5$Pb$_3$, maximized at temperatures between 100 K and 200 K, indicative of a non-coplanar spin texture in the magnetic state between $T_1$ and $T_2$. This is also supported by the observation of large magnetic anisotropy, which is only observed in a few other Gd-based compounds with non-coplanar spin textures. Our work thus provides a solid basis to understand the complex magnetic transitions in Gd$_5$Pb$_3$. The THE in Gd$_5$Pb$_3$ occurs at temperatures much higher than those of other rare earth compounds. Furthermore, our work opens the possibility for realizing THE in other such systems, as several of the compounds in the cluster near Gd$_5$Pb$_3$ have been reported to feature multiple magnetic transitions that warrant extensive magnetotransport studies on single crystals. We can probe the validity of our strategy in either non-centrosymmetric compounds with DMI (Gd$_4$Bi$_3$) or square-net compounds with higher order coupling (GdScSi or GdScGe) (steps 3-4).
\bibliography{References}
\clearpage
\onecolumngrid

\begingroup

\setcounter{figure}{0}
\setcounter{table}{0}
\renewcommand{\thefigure}{S\arabic{figure}}
\renewcommand{\thetable}{S\arabic{table}}
\section*{Supplementary Information}
\section*{This supplementary information includes:}
Supplementary Note 1: Structural characterization

Supplementary Note 2: Ordering temperatures and shortest Gd-Gd distances

Supplementary Note 3: Magnetization of \texorpdfstring{$\mathrm{Gd_5}\mathrm{Pb_3}$}{Gd5Pb3}

Supplementary Note 4: Specific heat of \texorpdfstring{$\mathrm{Gd_5}\mathrm{Pb_3}$}{Gd5Pb3} and non-magnetic analog \texorpdfstring{$\mathrm{La_5}\mathrm{Pb_3}$}{La5Pb3}

Supplementary Note 5: Temperature-dependent Raman scattering of \texorpdfstring{$\mathrm{Gd_5}\mathrm{Pb_3}$}{Gd5Pb3}

Supplementary Note 6: Resistivity of \texorpdfstring{$\mathrm{La_5}\mathrm{Pb_3}$}{La5Pb3}

Supplementary Note 7: Magnetoresistance of \texorpdfstring{$\mathrm{Gd_5}\mathrm{Pb_3}$}{Gd5Pb3}

Supplementary Note 8: Hall resistivity of \texorpdfstring{$\mathrm{Gd_5}\mathrm{Pb_3}$}{Gd5Pb3}

Supplementary Note 9: Phase diagram of \texorpdfstring{$\mathrm{Gd_5}\mathrm{Pb_3}$}{Gd5Pb3}

\section{Structural characterization}

We performed powder X-ray diffraction measurements on Gd$_5$Pb$_3$ in an air-tight sample holder at 300 K. The diffraction peaks along with the Rietveld refinement is shown in Fig. \ref{pxrd_refinement}. The fitting statistics were $\chi^2 = 2.58$, R$_{wp} = 6.62\%$ and R$_{exp} = 4.12\%$. The lattice parameters, interatomic distances and atomic positions are given in supplementary tables \ref{Crystallographic_parameters} and \ref{Atomic_positions}.

\begin{table}[H]
    \centering
    \caption{Crystallographic parameters for \texorpdfstring{$\mathrm{Gd_5}\mathrm{Pb_3}$}{Gd5Pb3} at 300 K}
    \begin{tabular}{| c | c |}
        \hline
            Formula & Gd$_5$Pb$_3$ \\
            \hline
            Space group & $P6_3/mcm$ (No. 193)\\
            \hline
            $a$ (\AA) & 9.0526(4)\\
            \hline
            $c$ (\AA) & 6.6041(3) \\
            \hline
            $V$ (\AA$^3$) & 468.70(4) \\
            \hline
            $Z$ & 2 \\
            \hline
            d\textsubscript{Gd1-Gd1}, out-of-plane (\AA) & 3.3020(2) \\
            \hline
            d\textsubscript{Gd1-Gd1}, in-plane (\AA) & 5.2266(3) \\
            \hline
            d\textsubscript{Gd2-Gd2}, out-of-plane (\AA) & 3.9527(1) \\
            \hline
            d\textsubscript{Gd2-Gd2}, in-plane (\AA) & 3.7631(2) \\
            \hline
            d\textsubscript{Gd1-Gd2} (\AA) & 3.8852(2) \\
            \hline
    \end{tabular}
    \label{Crystallographic_parameters}
\end{table}%

\begin{table}[H]
    \centering
    \caption{Atomic positions for \texorpdfstring{$\mathrm{Gd_5}\mathrm{Pb_3}$}{Gd5Pb3} }
    \begin{tabular}{|c|c|c|c|c|c|}
        \hline
            Atom & Wyckoff & $x$ & $y$ & $z$ & Occupancy\\
            \hline
            Gd1 & $4d$ & $\frac{1}{3}$ & $\frac{2}{3}$ & 0.00000 & 1\\
            \hline
            Gd2 & $6g$ & $0.24000$ & $0.00000$ & 0.25000 & 1\\
            \hline
            Pb & $6g$ & $0.60000$ & $0.00000$ & 0.25000 & 1\\
            \hline
    \end{tabular}
    \label{Atomic_positions}
\end{table}%

 \begin{figure}[H]
    \centering
    \includegraphics[width=0.6\linewidth]{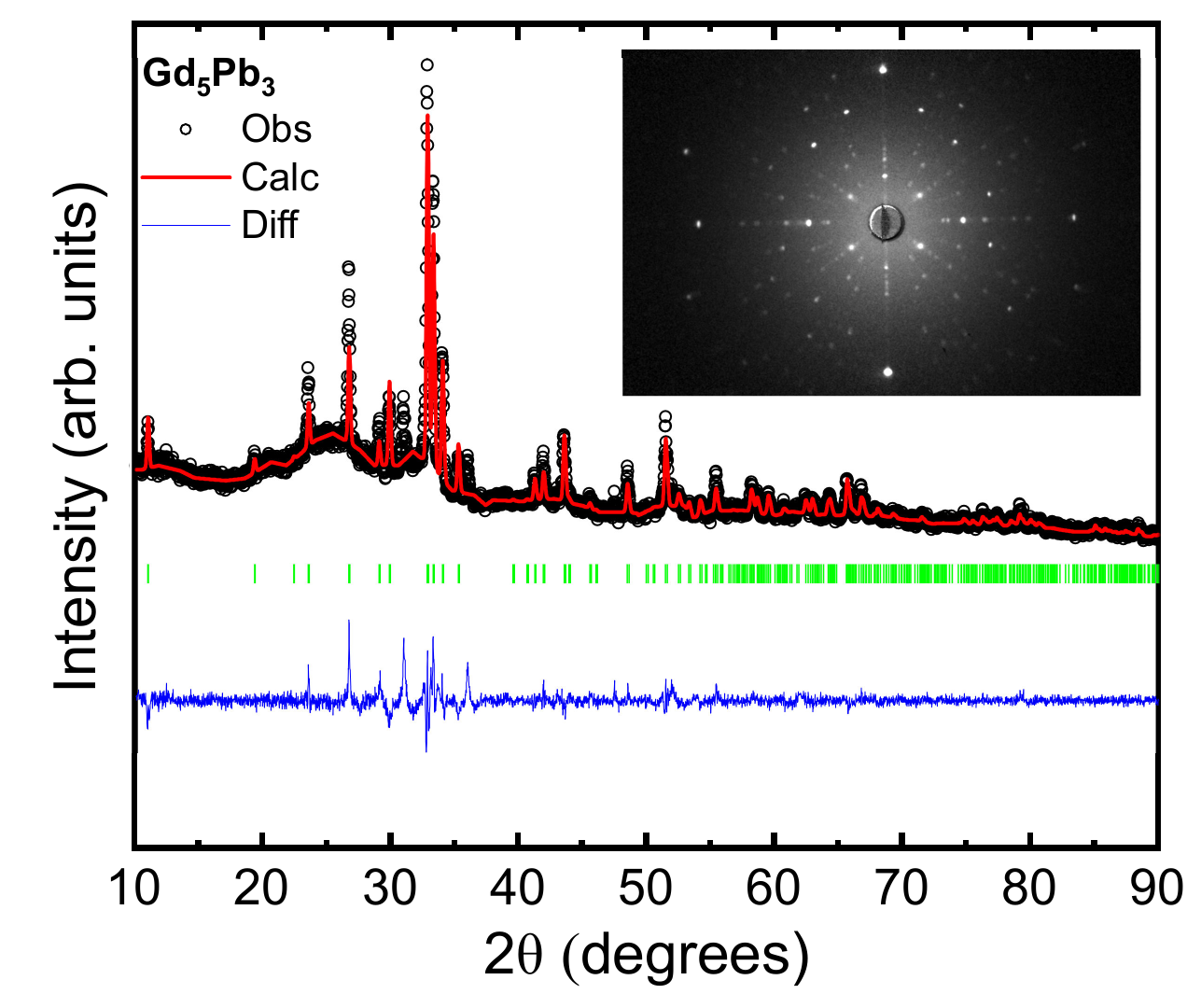}
    \caption{Powder x-ray diffraction pattern of Gd$_5$Pb$_3$ (black circles) taken at T = 300 K along with the Rietveld refinement (red line). The blue line represents the difference between the measured and fitted intensity. The green ticks correspond to the Bragg peak positions. The inset shows X-ray Laue diffraction pattern taken with the X-ray beam along the c-axis.}
    \label{pxrd_refinement}
\end{figure}

 \begin{figure}[H]
    \centering
    \includegraphics[width=1.0\linewidth]{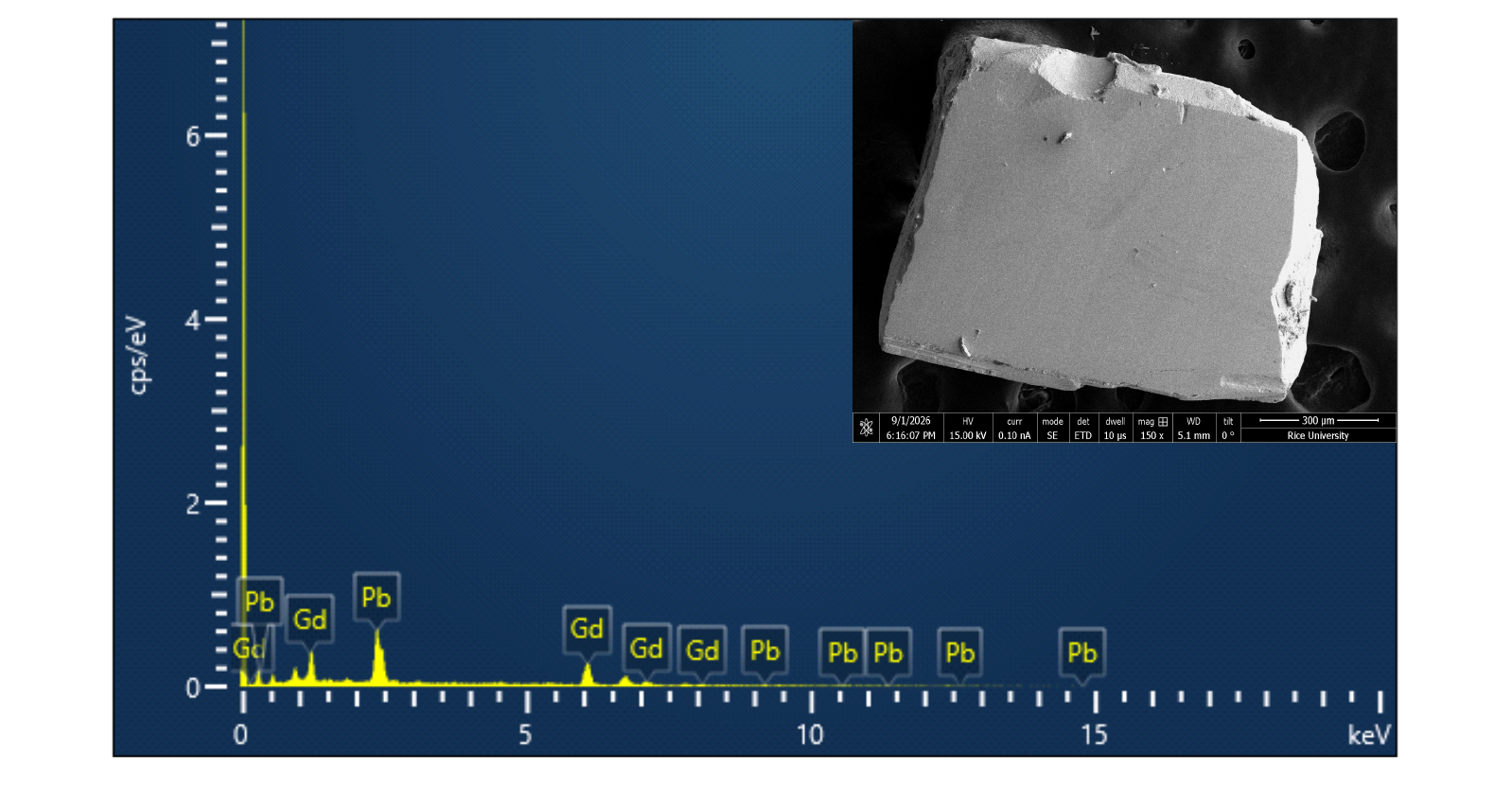}
    \caption{Energy dispersive X-ray spectroscopy (EDX) pattern of a Gd$_5$Pb$_3$ single crystal. Only Gd and Pb elements can be identified. The inset shows a scanning electron microscopy (SEM) image of a single crystal.}
    \label{edx}
\end{figure}

\begin{table}[H]
    \centering
    \caption{The atomic percentages of Gd and Pb determined from EDX measurements for two crystals. The average stoichiometry of each crystal is determined by examining 5 different locations on the sample surface. The average compositions of the two crystals are very close to the chemical formula of Gd$_5$Pb$_3$.}
    \begin{tabular}{|c|c|c|c|c|c|c|}
        \hline
            Sample No. \# & Position 1 & Position 2 & Position 3 & Position 4 & Position 5 & Average Stoichiometry\\
            \hline
            1 & \makecell[c]{Gd: 61.8\% \\ Pb: 38.2\%} &\makecell[c]{Gd: 59.8\% \\ Pb: 40.2\%}   &\makecell[c]{Gd: 65.0\% \\ Pb: 35.0\%}   & \makecell[c]{Gd: 61.6\% \\ Pb: 38.4\%} & \makecell[c]{Gd: 59.4\% \\ Pb: 40.6\%} &Gd$_{4.92}$Pb$_{3.08}$\\
            \hline
            2 &  \makecell[c]{Gd: 64.6\% \\ Pb: 35.4\%}& \makecell[c]{Gd: 58.9\% \\ Pb: 41.1\%} & \makecell[c]{Gd: 71.9\% \\ Pb:28.1\%} & \makecell[c]{Gd: 57.1\% \\ Pb: 42.9\%}& \makecell[c]{Gd: 64.8\% \\ Pb: 36.1\%}& Gd$_{5.07}$Pb$_{2.93}$\\
            \hline
    \end{tabular}
    \label{edx_values}
\end{table}%

\section{Ordering temperatures and shortest \texorpdfstring{$\mathbf{Gd-Gd}$}{Gd-Gd} distances}

Supplementary table \ref{tord_d} lists the ordering temperatures ($T_C$ is the Curie temperature, $T_N$ is the N\'eel temperature) and the shortest Gd-Gd distances in 50 ferromagnetic (columns 1 - 4) and 50 antiferromagnetic (columns 5 - 8) compounds with Gd-only magnetic moments. The distances are obtained from the Inorganic Crystal Structure Database (ICSD) \cite{zagoracRecentDevelopmentsInorganic2019}. 

\begin{longtable}{|c|c|c|c||c|c|c|c|}
 \caption{List of magnetic Gd compounds}
  \label{tord_d} \\
 \hline
 Compound & $T_C$ (K) & $d_{\text{Gd-Gd}}$ (\AA) & Ref. &Compound & $T_N$ (K) & $d_{\text{Gd-Gd}}$ (\AA) & Ref. \\
 \endfirsthead
\hline
 \cellcolor{gray!50}GdScSi & \cellcolor{gray!50} 354 & \cellcolor{gray!50}3.766 & \cellcolor{gray!50}\cite{couillaudMagnetocaloricPropertiesGdScSi2011a} & GdCu & 150 & 3.501 & \cite{postnikovElectronicMagneticStructure1992} \\
\hline
\cellcolor{gray!50}GdScGe & \cellcolor{gray!50}350 & \cellcolor{gray!50}3.7716 &\cellcolor{gray!50} \cite{manfrinettiSingleCrystalStudy2002} & Gd$_2$MgGe$_2$ & 150 & 3.7959 & \cite{choeCrystalStructureMagnetism2001} \\ 
\hline
\cellcolor{gray!50}Gd$_5$Si$_4$ & \cellcolor{gray!50}336 &\cellcolor{gray!50} 3.43748 & \cellcolor{gray!50}\cite{pecharskyGiantMagnetocaloricEffect1997} & GdAg & 132.8 & 3.672 & \cite{chattopadhyayAntiferromagneticPhaseTransition1996} \\ 
\hline
\cellcolor{gray!50}Gd$_4$Bi$_3$ & \cellcolor{gray!50}335 & \cellcolor{gray!50}3.61479 & \cellcolor{gray!50}\cite{szadeMagnetismElectronicStructure2000} & Gd$_5$Ge$_4$ & 128 & 3.57 & \cite{ouyangMagneticAnisotropyMagnetic2006} \\ 
\hline
\cellcolor{gray!50}GdTiSi & \cellcolor{gray!50}294 & \cellcolor{gray!50}3.63 & \cellcolor{gray!50}\cite{skorekElectronicStructureMagnetism2001} & Gd$_3$Rh & 112 & 3.51855 & \cite{monteiroPhysicalPropertiesGd3Ru2015} \\ 
\hline
\cellcolor{gray!50}Gd &\cellcolor{gray!50} 293 & \cellcolor{gray!50}3.573 & \cellcolor{gray!50} \cite{cableNeutronDiffractionStudy1968} & Gd$_2$RuGe$_2$ & 97.2 & 3.5717 & \cite{bouletStructuralChemistryMagnetic2000} \\ 
\hline
\cellcolor{gray!50}Gd$_5$Pb$_3$ & \cellcolor{gray!50}285 &\cellcolor{gray!50} 3.30205 &\cellcolor{gray!50} \cite{marcinkovaStrongMagneticCoupling2015}, This Work & GdRh$_2$Ge$_2$ & 93 & 4.127 & \cite{gignouxMagneticPropertiesRRh2Ge22000} \\ 
\hline
\cellcolor{gray!50}Gd$_3$Al$_2$ & \cellcolor{gray!50}281 &\cellcolor{gray!50} 3.5564 &\cellcolor{gray!50} \cite{pecharskyMagnetocaloricPropertiesGd3Al22002} & GdBPt$_2$ & 87 & 3.75191 & \cite{satoNewGdbasedMagnetic2022} \\ 
\hline
\cellcolor{gray!50}GdZn & \cellcolor{gray!50}270 & \cellcolor{gray!50}3.618 &\cellcolor{gray!50} \cite{zhaoLargeAnomalousHall2021} & Gd$_5$Ge$_3$ & 82 & 3.2201 & \cite{rogerMn5Si3typeHostinterstitialBoron2006} \\ 
\hline
\cellcolor{gray!50}GdCd & \cellcolor{gray!50}265 & \cellcolor{gray!50}3.7501 &\cellcolor{gray!50} \cite{zhaoLargeAnomalousHall2021} & GdSe & 60 & 4.08142 & \cite{rotterDipoleInteractionMagnetic2003} \\ 
\hline
\cellcolor{gray!50}Gd$_5$Sb$_3$ & \cellcolor{gray!50}265 & \cellcolor{gray!50}3.16 & \cellcolor{gray!50}\cite{samathamMagnetizationResistivitySpecific2018} & Gd$_5$Sn$_3$ & 60 & 3.297 & \cite{rogerMn5Si3typeHostinterstitialBoron2006} \\ 
\hline
Gd$_2$In & 187 & 3.371 & \cite{mcalisterMagneticElectricalProperties1984} & Gd$_5$Si$_3$ & 55 & 3.2103 & \cite{rogerMn5Si3typeHostinterstitialBoron2006} \\ 
\hline
GdAl$_2$ & 170 & 3.41647 & \cite{oishiNegativeVolumeMagnetostriction2010} & Gd$_3$Ru & 54 & 3.51065 & \cite{monteiroPhysicalPropertiesGd3Ru2015} \\ 
\hline
Gd$_5$CoSi$_2$ & 168 & 3.362 & \cite{mayerNewTernarySilicide2011a} & GdS & 50 & 3.93505 & \cite{rotterDipoleInteractionMagnetic2003} \\ 
\hline
Gd$_3$Co$_2$Ge$_4$ & 135 & 3.3984 & \cite{xiaoSuccessiveMagneticTransitions2025} & GdAu$_2$ & 50 & 3.728 & \cite{rotterDipoleInteractionMagnetic2003} \\ 
\hline
GdFeSi & 125 & 3.7011 & \cite{nikitinMagneticAnisotropyMagnetic1998} & GdRu$_2$Si$_2$ & 47 & 4.164 & \cite{garnierAnisotropicMetamagnetismGdRu2Si21995} \\ 
\hline
Gd$_5$PtSb$_2$ & 125 & 3.4 & \cite{morozkinNewTernaryYb5Sb3type2013a} & Gd$_8$Ag$_{19.5}$Al$_{45.2}$ & 46.5 & 5.783 & \cite{tyvanchukCrystalStructureMagnetic2019} \\ 
\hline
Gd$_2$CdCu$_2$ & 116.7 & 3.9451 & \cite{schappacherStructureMagneticProperties2009} & GdAgSn & 34 & 3.7163 & \cite{baranNeutronDiffractionStudy1997} \\ 
\hline
Gd$_5$PbBi$_2$ & 113 & 3.425 & \cite{morozkinNewTernaryYb5Sb3type2013a} & GdCoIn$_5$ & 30 & 4.567 & \cite{betancourthLowtemperatureMagneticProperties2015} \\ 
\hline
Gd$_5$Bi$_3$ & 112 & 3.2093 & \cite{svitlykGd5Ni096Sb204Gd5Ni071Bi229Crystal2008} & Gd$_3$Ge$_4$ & 29 & 3.7746 & \cite{tobashStructurePropertiesGd32007} \\ 
\hline
Gd$_5$Bi$_3$ & 110 & 3.39 & \cite{szadeMagnetismElectronicStructure2000} & GdP & 28 & 4.04677 & \cite{rotterDipoleInteractionMagnetic2003} \\ 
\hline
GdCoAl & 100 & 3.21413 & \cite{jaroszCrystallographicElectronicStructure2000} & GdBi & 27.5 & 4.46241 & \cite{dwariLargeUnsaturatedMagnetoresistance2023} \\ 
\hline
GdRh$_2$B$_2$C & 100 & 3.7491 & \cite{isidaPhysicalPropertiesNew2000} & Gd$_2$TiAl$_3$ & 26.1 & 3.4404 & \cite{giesselmannRE2TiAl3REGd2023} \\ 
\hline
GdRh$_3$B$_2$ & 93 & 3.115 & \cite{obirakiMagneticFermiSurface2006a} & GdBe$_{13}$ & 26 & 5.14 & \cite{besnusMagneticPropertiesSpecific1996} \\ 
\hline
Gd$_{11}$Ni$_4$In$_9$ & 87 & 3.468 & \cite{szytulaMagneticPropertiesSpecific2014} & GdCu$_5$ & 26 & 4.99217 & \cite{barandiaranMagneticPropertiesMagnetic1989} \\ 
\hline
GdBPt$_2$ & 87 & 3.75191 & \cite{satoNewGdbasedMagnetic2022} & GdCuSn & 24 & 4.5341 & \cite{rotterDipoleInteractionMagnetic2003} \\ 
\hline
Gd$_{11}$Ge$_8$In$_2$ & 84 & 3.391 & \cite{cheungStructureMagneticMagnetocaloric2011a} & Gd$_2$Cl$_3$ & 23 & 3.698 & \cite{kremerAntiferromagneticOrderingFrustrated2024} \\ 
\hline
GdCuAl & 83 & 3.6854 & \cite{jaroszCrystallographicElectronicStructure2000} & GdGa$_2$ & 22 & 4.22 & \cite{barandiaranMagneticPropertiesMagnetic1989} \\ 
\hline
GdCuMg & 82.2 & 3.907 & \cite{steinGdCuMgZrNiAltypeStructure2017} & Gd$_2$PdSi$_3$ & 21 & 4.0784 & \cite{sahaMagneticAnisotropyFirstorderlike1999} \\ 
\hline
Gd$_4$Rh$_9$Ga$_5$ & 78.1 & 3.412 & \cite{seidelTernaryGallidesRE42017} & GdNi$_2$B$_2$C & 20 & 3.575 & \cite{rotterDipoleInteractionMagnetic2003} \\ 
\hline
GdRuGe & 72 & 3.6187 & \cite{pencMagneticPropertiesRRuGe1998} & Gd$_3$Ru$_4$Al$_{12}$ & 18.6 & 3.7147 & \cite{nakamuraMagneticPhasesFrustrated2023} \\ 
\hline
Gd$_5$Si$_2$B$_8$ & 70 & 3.4963 & \cite{rogerCrystalStructuresPhysical2005} & GdAuGe & 17 & 3.709 & \cite{kurumajiSingleCrystalGrowths2023} \\ 
\hline
GdNi & 69 & 3.5873 & \cite{uhlirovaMagneticMagnetoelasticProperties2007} & GdPtPb & 15.5 & 3.9649 & \cite{manniGdPtPbNoncollinearAntiferromagnet2017} \\ 
\hline
GdPtIn & 67.5 & 3.94723 & \cite{morosanMagneticOrderingEffects2005} & Gd$_2$Te$_3$ & 15.3 & 4.263 & \cite{muthuselvamGd2Te3Antiferromagnetic2019} \\ 
\hline
GdPdCd & 62.5 & 3.9178 & \cite{hoffmannFerromagneticOrderingGdPdCd2002} & GdAs & 15.2 & 4.1394 & \cite{rotterDipoleInteractionMagnetic2003} \\ 
\hline
GdNiIn$_2$ & 60.5 & 3.8877 & \cite{latkaMagneticProperties155Gd2006} & GdPdSn & 14.5 & 3.6978 & \cite{heymannHighpressureHightemperatureCrystal2016} \\ 
\hline
GdNiAl & 60 & 3.63762 & \cite{jaroszCrystallographicElectronicStructure2000} & Gd$_2$Rh$_3$Al$_9$ & 13.6 & 4.1767 & \cite{hiroakiHighpressureSynthesisMagnetic2024} \\ 
\hline
Gd$_2$RhAl$_3$ & 51.3 & 3.28 & \cite{eustermannSynthesisPhysicalMagnetocaloric2024} & GdAgGe & 13 & 3.729 & \cite{gibsonTernaryGermanidesLnAgGe1996} \\ 
\hline
GdPdAl & 48 & 3.7391 & \cite{talikMagneticPropertiesGdPdAl2001} & GdSbTe & 12 & 5.2172 & \cite{sankarCrystalGrowthMagnetic2019} \\ 
\hline
Gd$_3$Pt$_{23}$Si$_{11}$ & 42 & 5.9485 & \cite{opagisteFerromagnetismNovelCompounds2014} & GdAu$_2$Si$_2$ & 12 & 4.244 & \cite{rotterDipoleInteractionMagnetic2003} \\ 
\hline
GdNi$_5$ & 32 & 3.96774 & \cite{rotterDipoleInteractionMagnetic2003} & Gd$_3$Au$_4$Ga$_7$ & 12 & 3.7871 & \cite{grinCrystalChemistryMagnetic1994} \\ 
\hline
Gd$_2$Ni$_2$Pb & 32 & 3.707 & \cite{gulayPhysicalPropertiesTernary2003} & Gd$_2$Ir$_3$Ge$_5$ & 11.9 & 4.107 & \cite{singhMagneticOrderingSuperconductivity2004} \\ 
\hline
Gd$_3$Ni$_8$Sn$_4$ & 24 & 3.7914 & \cite{romakaCrystallographicMagneticElectrical2017} & GdFe$_2$Ge$_2$ & 10.8 & 4.002 & \cite{avilaAnisotropicMagnetizationSpecific2004} \\ 
\hline
GdNi$_4$Ga & 20 & 3.715 & \cite{joshiMagneticPropertiesTernary2006} & GdNiNB & 10.3 & 5.594 & \cite{songSynthesisStructureProperties2024} \\ 
\hline
GdCo$_2$B$_2$C & 17.2 & 3.548 & \cite{zhangMagneticPropertiesMagnetocaloric2020} & GdInCu$_2$ & 10 & 4.6959 & \cite{rotterDipoleInteractionMagnetic2003} \\ 
\hline
GdNi$_2$Cd$_{20}$ & 13.8 & 6.7134 & \cite{burnettStructurePhysicalProperties2014} & GdPtBi & 9 & 4.72347 & \cite{morozkinMagneticMagnetocaloricProperties2024} \\ 
\hline
GdRh$_3$B & 12 & 4.182 & \cite{joshiMagneticBehaviorRh2009a} & Gd$_3$Co$_4$Ge$_{13}$ & 8.14 & 4.3913 & \cite{shekharAnomalousHallEffect2018} \\ 
\hline
GdCr$_2$Si$_2$C & 10.9 & 3.9957 & \cite{maStructuralMagneticCryogenic2022} & GdInCu$_4$ & 7 & 5.1103 & \cite{rotterDipoleInteractionMagnetic2003} \\ 
\hline
GdRe$_2$Al$_{10}$ & 7.2 & 4.0083 & \cite{sefatPropertiesRe22009a} & GdCdNi$_4$ & 4.5 & 4.9908 & \cite{leeSuppressionMolecularField2023} \\ 
\hline
GdRh$_3$C & 3.5 & 4.135 & \cite{joshiMagneticBehaviorRh2009a} & GdBa$_2$Cu$_3$O$_7$ & 2.2 & 3.8 & \cite{rotterDipoleInteractionMagnetic2003} \\
\hline
\end{longtable}

\section{Magnetization of \texorpdfstring{$\mathbf{Gd_5}\mathbf{Pb_3}$}{Gd5Pb3}}

The isothermal magnetization curves up to 7 T are shown in Fig. \ref{lowfieldMH}. The high magnetic field  magnetization isotherms M(H) up to 60 T are shown in Fig. \ref{highfieldMH}. The critical fields for the metamagnetic transitions are determined from the peaks in the derivative $\text{d}M/ \text{d}H$. 

\begin{figure}[H]
    \centering
    \includegraphics[width=1.0\linewidth]{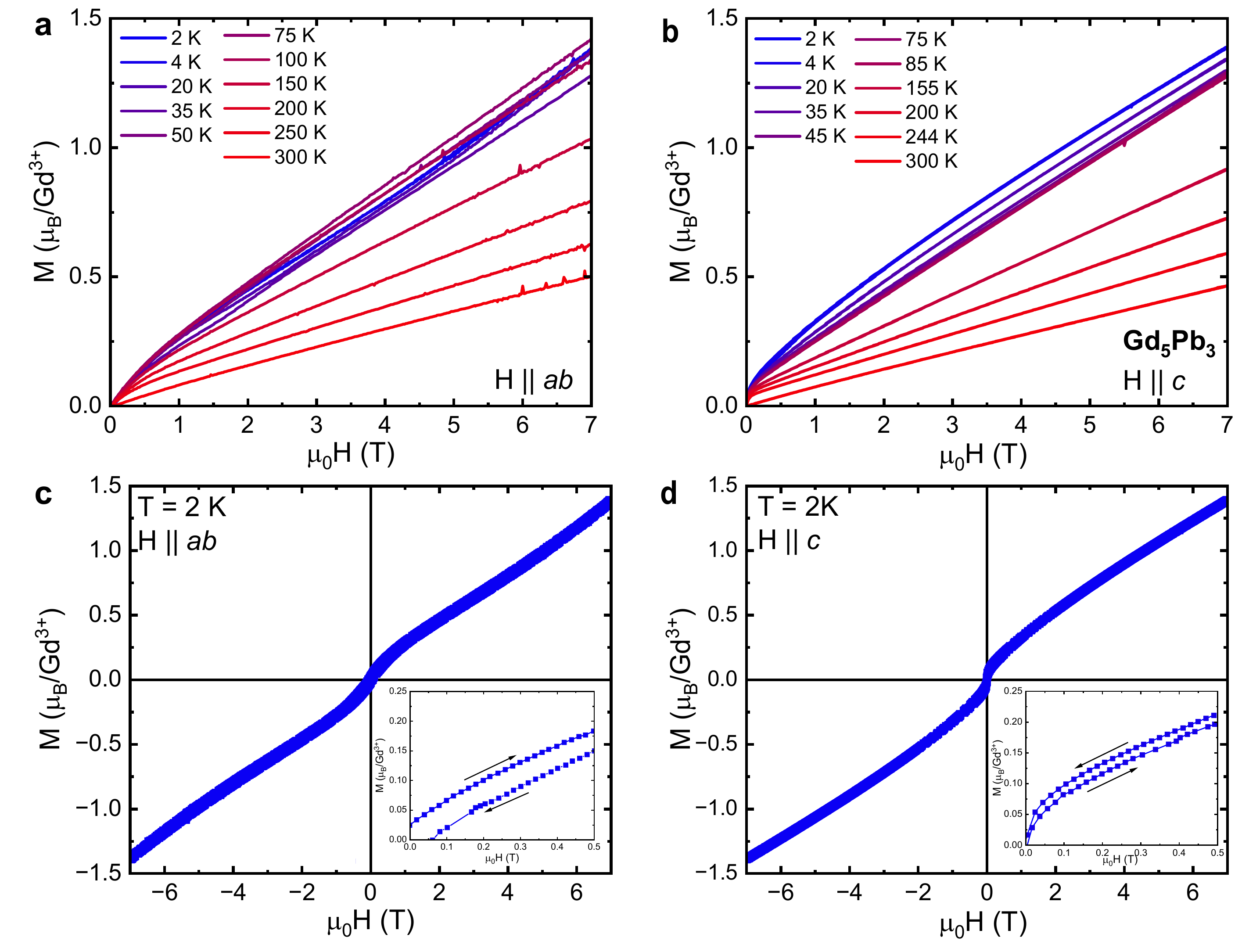}
    \caption{ (\textbf{a}, \textbf{b}) Magnetization ($M$) as a function of magnetic field ($\mu_0H$) up to $\mu_0H = 7$ T for field applied parallel to the (\textbf{a}) $ab$-plane and (\textbf{b}) $c$-axis. The temperatures range from 2 K to 300 K. (\textbf{c}, \textbf{d}) Four quadrant $M(H)$ at 2 K for (\textbf{c}) $H || ab$ and (\textbf{d}) $H || c$. The inset is the low field $M(H)$ showing very little hystersis.}
    \label{lowfieldMH}
\end{figure}

 \begin{figure}[H]
    \centering
    \includegraphics[width=1.0\linewidth]{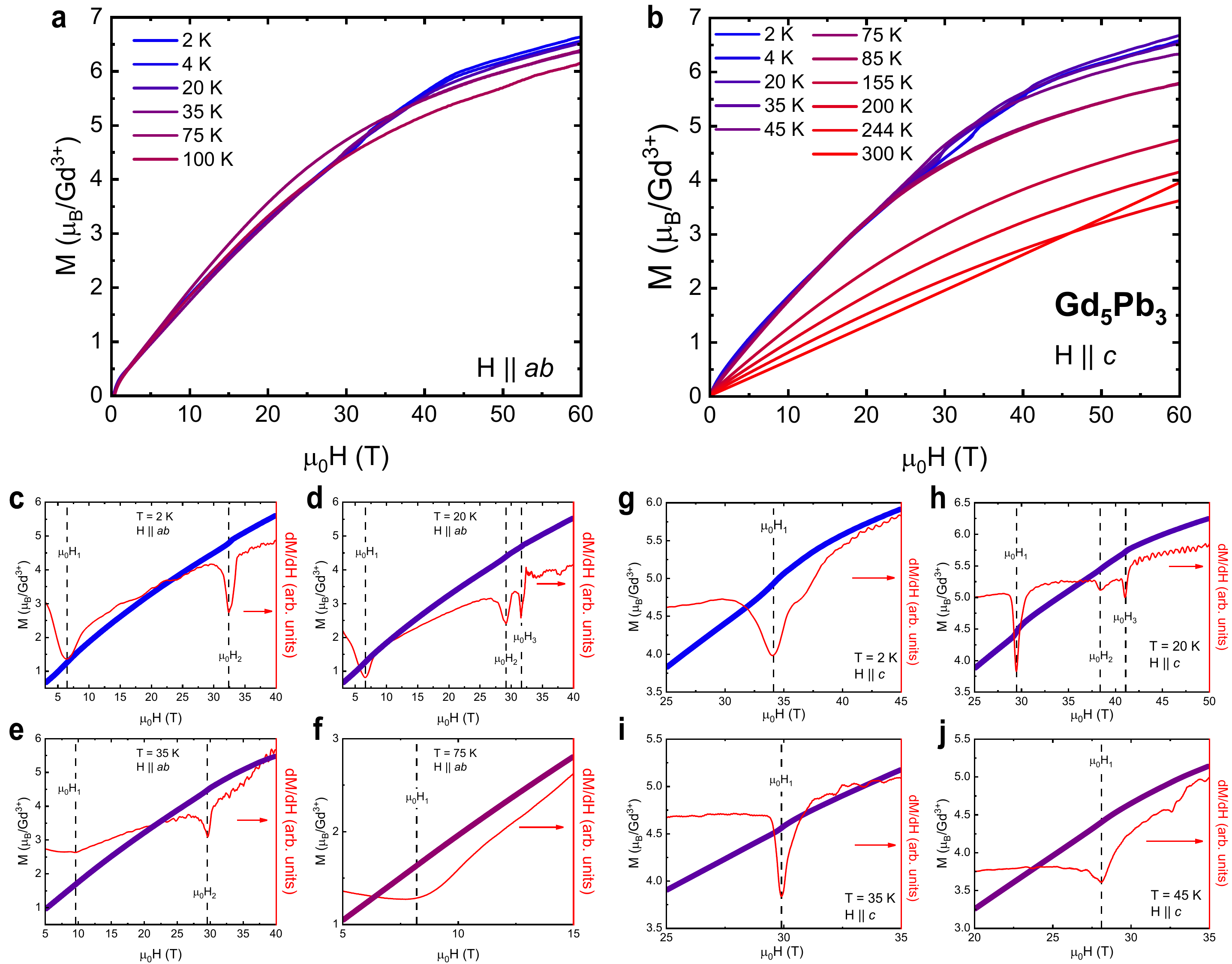}
    \caption{(\textbf{a}, \textbf{b}) Magnetization ($M$) as a function of magnetic field ($\mu_0H$) up to $\mu_0H = 60$ T for field applied parallel to the (\textbf{a}) $ab$-plane and (\textbf{b}) $c$-axis. The temperatures range from 2 K to 300 K. (\textbf{c} - \textbf{j}) $M(H)$ curves (left axis) displaying metamagnetic transitions and derivatives ($\text{d}M/ \text{d}H$) (red, right axis) with peaks corresponding to critical fields for field parallel to the $ab$-plane at (\textbf{c}) 2 K, (\textbf{d}) 20 K, (\textbf{e}) 35 K, (\textbf{f}) 75 K and for field parallel to the $c$-axis at (\textbf{g}) 2 K, (\textbf{h}) 20 K, (\textbf{i}) 35 K, and (\textbf{j}) 45 K.}
    \label{highfieldMH}
\end{figure}
The magnetic susceptibility measured between 2 K and 300 K at various magnetic fields parallel to the $c$ axis and the $ab$ plane from 0.01 T to 7 T is shown in Fig. \ref{susceptibility}. The zero-field-cooled (ZFC) and field-cooled (FC) curves separate below $T_C$ for both directions. The transition temperatures $T_C$, $T_1$, $T_2$, $T_3$ are determined from peaks in the derivative $\text{d}M/ \text{d}T$. 

 \begin{figure}[H]
    \centering
    \includegraphics[width=1.0\linewidth]{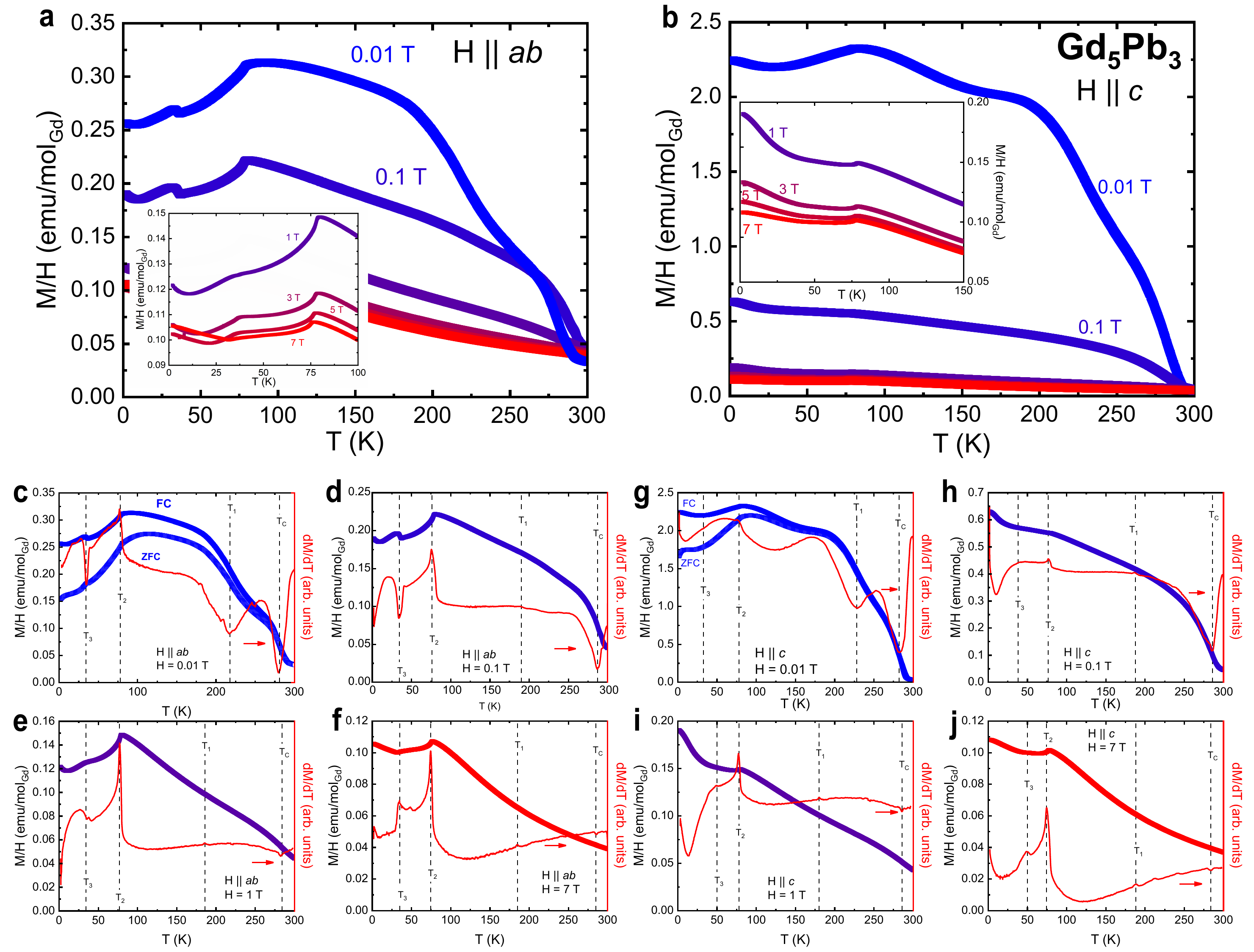}
    \caption{(\textbf{a}, \textbf{b}) Magnetic Susceptibility ($M/H$) as a function of temperature ($T$) up to $T = 300$ K for field applied parallel to the (\textbf{a}) $ab$-plane and (\textbf{b}) $c$-axis. The magnetic field ranges from 0.01 T to 7 T. The inset is a zoomed-in view of the high field $M/H$. (\textbf{c} - \textbf{j}) $M/H$ curves (left axis) and derivatives ($\text{d}M/ \text{d}T$)  (red, right axis) with peaks corresponding to the transition temperatures for field parallel to the $ab$-plane at (\textbf{c}) 0.01 T, (\textbf{d}) 0.1 T, (\textbf{e}) 1 T, (\textbf{f}) 7 T and for field parallel to the $c$-axis at (\textbf{g}) 0.01 T, (\textbf{h}) 0.1 T, (\textbf{i}) 1 T, and (\textbf{j}) 7 T. The ZFC and FC susceptibility curves are shown for 0.01 T for both directions.}
    \label{susceptibility}
\end{figure}

\section{Specific heat of \texorpdfstring{$\mathbf{Gd_5}\mathbf{Pb_3}$}{Gd5Pb3} and non-magnetic analog \texorpdfstring{$\mathbf{La_5}\mathbf{Pb_3}$}{La5Pb3}}

 \begin{figure}[H]
    \centering
    \includegraphics[width=1.0\linewidth]{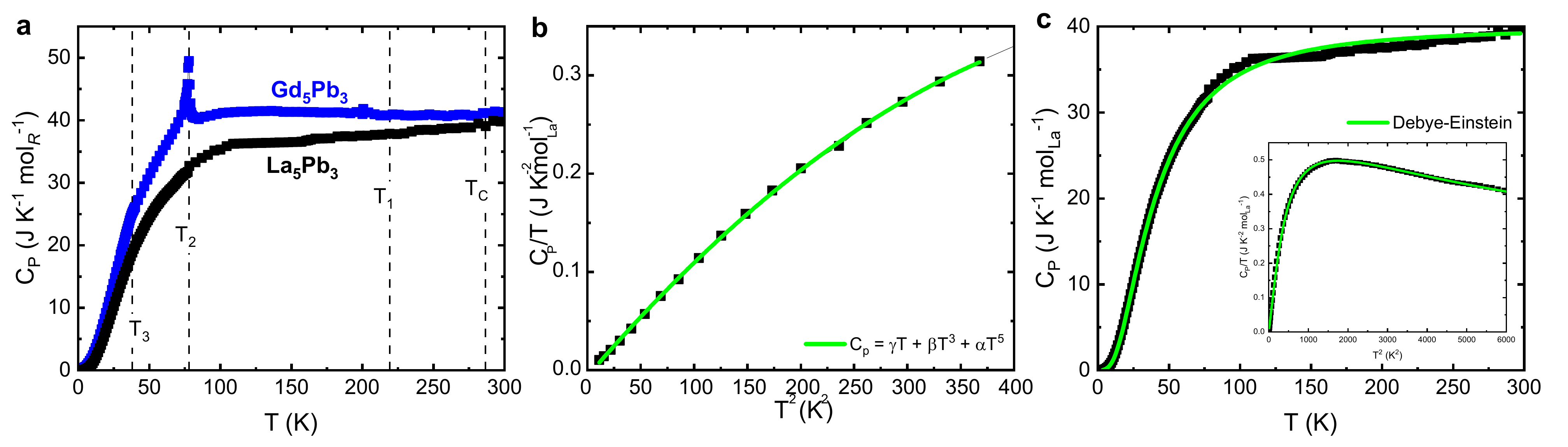}
    \caption{(a) Specific heat of Gd$_5$Pb$_3$ (blue) and La$_5$Pb$_3$ (black) measured from 1.8 K to 300 K. (b) Polynomial fitting of $C_P/T$ versus $T^2$ up to $T = 20$ K for La$_5$Pb$_3$ (c) Debye-Einstein fit for La$_5$Pb$_3$. The inset presents a zoomed-in view of $C_P/T$ versus $T^2$.}
    \label{heat_capacity}
\end{figure}

The heat capacity of Gd$_5$Pb$_3$ and La$_5$Pb$_3$ measured from 1.8 K to 300 K is shown in Fig. \ref{heat_capacity}. The electronic and phonon contribution below 20 K for La$_5$Pb$_3$ can be estimated using a simple polynomial equation $C_p = \gamma T + \beta T^3 + \alpha T^5$, which yields a small Sommerfeld coefficient ($\gamma$) $\approx 6$ mJ K$^{-2} $mol$^{-1}_{\text{La}}$,  $\beta \approx 1.26$ mJ K$^{-4}$ mol$^{-1}_{\text{La}}$ and $\alpha \approx -0.001$ mJ K$^{-6} $mol$^{-1}_{\text{La}}$. The Debye temperature ($\Theta_D$) is then calculated as 

\begin{equation}
    \Theta_D = \left( \frac{12 \pi^4 R}{5 \beta} \right)^{1/3}, 
\end{equation}
where $R$ is the universal gas constant. Therefore, $\Theta_D$ is approximately 115 K for La$_5$Pb$_3$ and 111 K for Gd$_5$Pb$_3$. A more accurate fitting for the phonon contribution can be performed using the Debye-Einstein model \cite{xinPropertiesCaB6Single2011, debnathElectronicCorrelationEffects2023, gamsjagerLowTemperatureHeat2018}, given by Eq. \ref{debye_einstein},
\begin{equation}\label{debye_einstein}
    C_P = mC_\text{Debye} + nC_\text{Einstein},
\end{equation}
where $C_\text{Debye}$ and $C_\text{Einstein}$ are given by
\begin{equation}
    C_\text{Debye} = 9R \left( \frac{T}{\Theta_D}\right)^3\int^{\frac{\Theta_D}{T}}_0 \frac{x^4 \text{e}^x}{(\text{e}^x - 1)^2}\text{d}x
\end{equation} 

\begin{equation}
    C_\text{Einstein} = 3R\left( \frac{\Theta_E}{T}\right)^2\frac{\text{e}^\frac{\Theta_E}{T}}{\left(\text{e}^\frac{\Theta_E}{T} - 1\right)^2}
\end{equation}
and $\Theta_E$ is the Einstein temperature, and $m+n$ should approximate the number of atoms in the formula unit. With this model, we determine a $\Theta_D$ of 134 K (130 K) and $\Theta_E$ of 217 K (210 K) for La$_5$Pb$_3$ (Gd$_5$Pb$_3$), with $m+n \approx 8$. 

\section{Temperature-dependent Raman scattering of \texorpdfstring{$\mathbf{Gd_5}\mathbf{Pb_3}$}{Gd5Pb3}}

 \begin{figure}[H]
    \centering
    \includegraphics[width=1.0\linewidth]{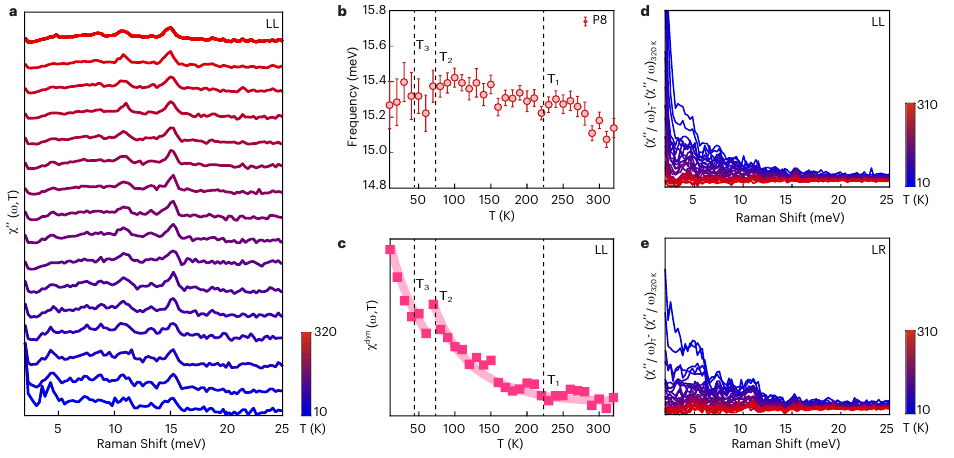}
    \caption{a-e, Temperature evolution of the Raman susceptibility, $\chi''$ ($\omega$,$T$) of Gd$_5$Pb$_3$ in the crossed (LR) circular polarization channel from 10 K to 320 K (a). Temperature dependence of the energy of phonon mode at 15.2 meV (P8) (b), and the dynamic Raman susceptibility, $\chi^{\text{dyn}}$($\omega$,$T$) (c). The vertical dished lines donate the magnetic transition temperatures. Temperature dependence of the Raman conductivity, $\chi''$ /$\omega$ in LL (d) and LR (e) circular polarization channel, where all the spectra are subtracted by the  $\chi''$ /$\omega$ at 320 K, and L (R) corresponds left- (right-) circularly polarized light. The color bar represents the temperature.}
    \label{Fig6}
\end{figure}

\section{Resistivity of \texorpdfstring{$\mathbf{La_5}\mathbf{Pb_3}$}{La5Pb3}}

 \begin{figure}[H]
    \centering
    \includegraphics[width=0.5\linewidth]{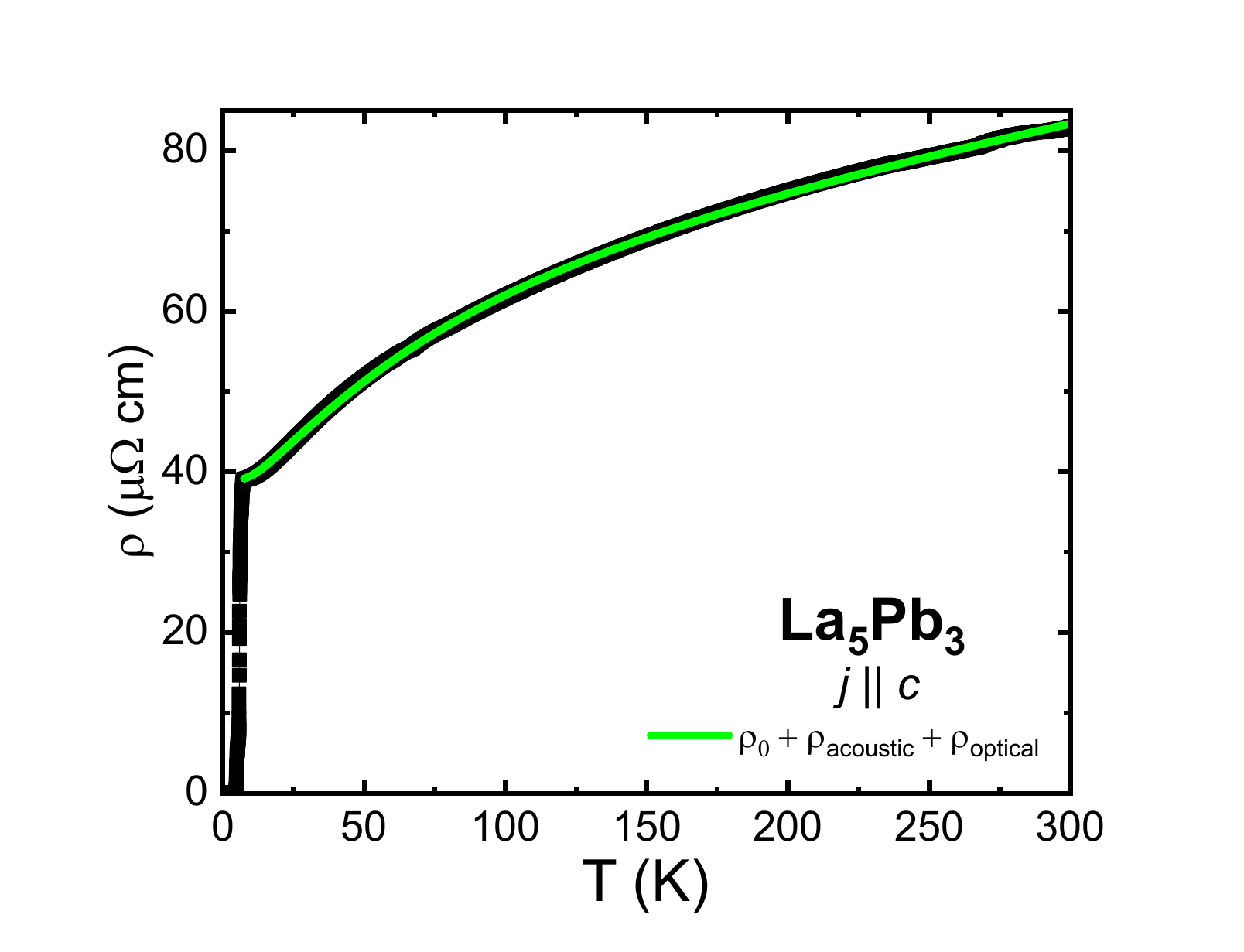}
    \caption{Resistivty of La$_5$Pb$_3$ as function of temperature. The fit (green) is performed using a Bloch-Gr\"{u}neisen function that takes into account electron scattering from acoustic and optical phonons.}
    \label{Fig7}
\end{figure}

The resistivity of La$_5$Pb$_3$ was measured from 1.8 K to 300 K. A small drop is associated with the superconductivity from remnant Pb flux on the surface of the sample. The resistivity has a downward turn at higher temperatures, and this behavior can be accurately captured using the sum of contributions resulting from the scattering of electrons with both acoustic and optical phonons \cite{sahakyanPhysicalPropertiesElectronic2016} using Eq. \ref{rho},
\begin{equation} \label{rho}
    \rho (T) = \rho_0 + \rho_{\text{acoustic}} + \rho_\text{optical},
\end{equation}
where $\rho_\text{acoustic}$ is given by the Bloch-Gr\"{u}neisen function for electron-acoustic phonon scattering:
\begin{equation}
    \rho_\text{acoustic}= C_a\left(\frac{T}{\Theta_D}\right)^5 \int^{\frac{\Theta_D}{T}}_0 \frac{x^5 \text{e}^x}{\left(\text{e}^x - 1 \right)^2}\text{d}x ,
\end{equation}
and $\rho_\text{optical}$ is given by:
\begin{equation}
    \rho_\text{optical}= C_o\left( \frac{\Theta_E}{T}\right)\frac{1}{\left(\text{e}^\frac{\Theta_E}{T} - 1 \right)^2}.
\end{equation}
Here, $\rho_0$ is the $T-$independent resistivity, and  $C_a$ and $C_o$ are numerical constants. From our fit, we obtained $\Theta_D \approx 64$ K and $\Theta_E \approx 158$ K for $C_a \approx 70$ and $C_o \approx -34$.

\section{Magnetoresistance of \texorpdfstring{$\mathbf{Gd_5}\mathbf{Pb_3}$}{Gd5Pb3}}

The transverse magnetoresistance MR (Fig. \ref{magnetoresistance}) is positive for temperatures below $T_2$, and negative for temperatures above $T_2$, which hints at the coupling of charge carriers and the magnetic structure. At high temperatures, the behaviour is typical of a ferromagnetic metal as the magnetic field suppresses spin-related scattering, leading to a decrease in resistance under field. At lower temperatures, the MR changes sign as the Lorentz force induced effects on the charge carriers dominate \cite{roychowdhuryGiantRoomTemperatureTopological2024, polashSpinFluctuationsYield2021}. The MR has a $H^2$ dependence until 3.5 T for temperatures below $T_3$, and until 2 T for temperatures below $T_2$, and it peaks at 10\% \% at 10 K around 7 T before leveling off. The critical field at which the MR peaks aligns with the critical field for the first metamagnetic transition determined from the magnetization isotherms. 

 \begin{figure}[H]
    \centering
    \includegraphics[width=1.0\linewidth]{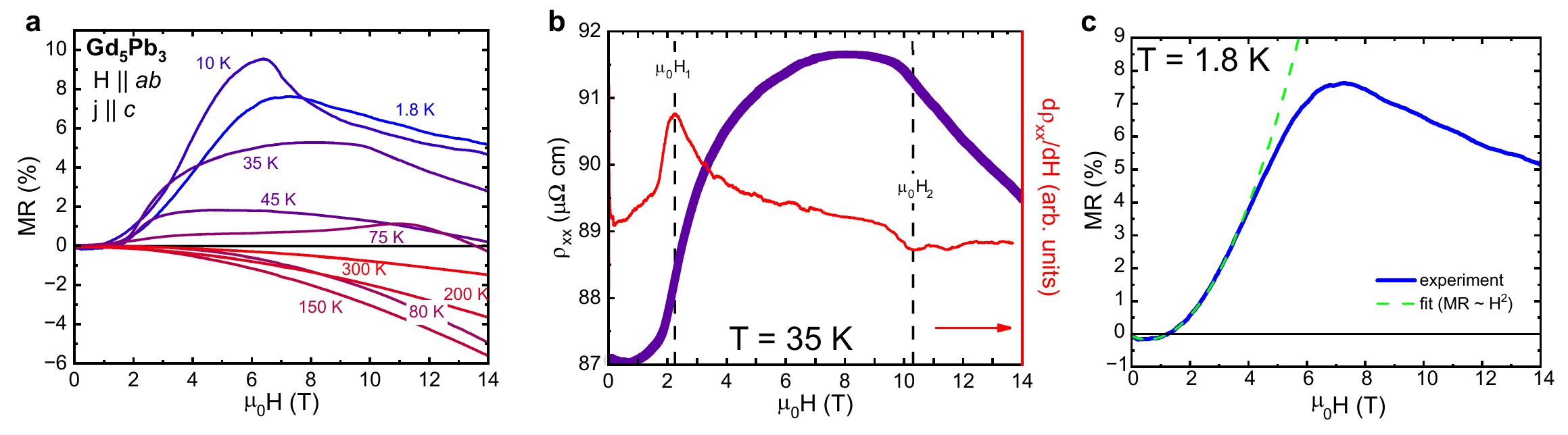}
    \caption{(a) Magnetoresistance (MR) of Gd$_5$Pb$_3$ as a function of field at different temperatures. (b) Metamagnetic transitions occuring in Gd$_5$Pb$_3$. The critical fields are determined by peaks in d$\rho_{xx}$/dH. (c) $H^2$ dependence of MR}
    \label{magnetoresistance}
\end{figure}

\section{Topological Hall effect in \texorpdfstring{$\mathbf{Gd_5}\mathbf{Pb_3}$}{Gd5Pb3}}
 \begin{figure}[H]
    \centering
    \includegraphics[width=1.0\linewidth]{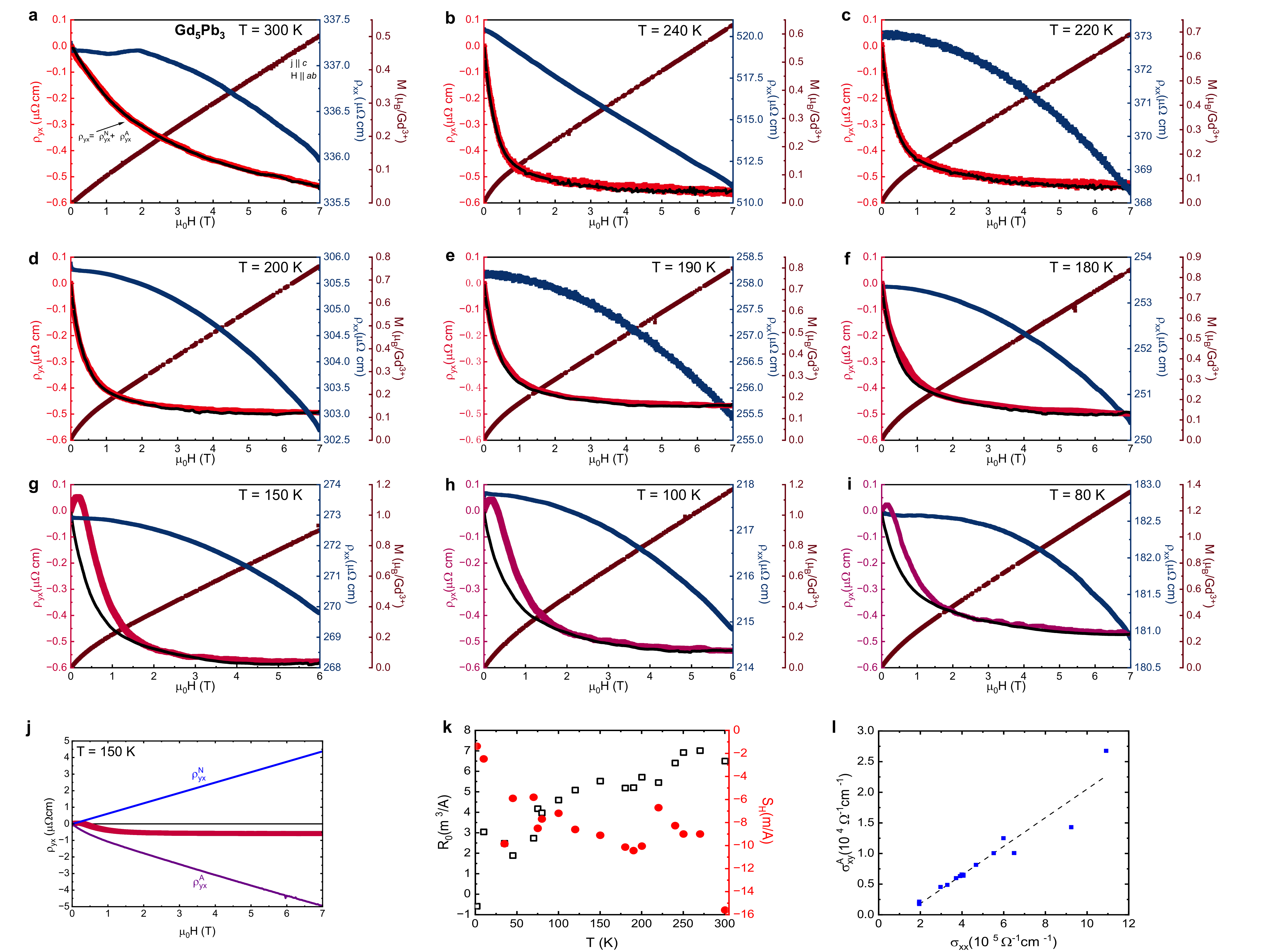}
    \caption{(a) - (i) Hall resistivity and fits at various temperatures (j) Normal (blue) and Anomalous (purple) contributions to the Hall resistivity for T = 150 K (h) Normal Hall coefficient ($R_0$) and Anomalous Hall coefficient ($S_H$) for various temperatures (i) Anomalous Hall conductivity ($\sigma^A_{xy}$) versus longitudinal conductivity ($\sigma_{xx}$)}
    \label{fitting values}
\end{figure}

We achieved the best fits to Hall resistivity data after considering a two-band model consisting of electron and hole carriers to express the normal Hall resistivity \cite{leiHighMobilityVan2020}, where $R_0$ is given by Eq. \ref{twoband} and a dominant skew scattering anomalous Hall resistivity mechanism. 

\begin{equation}\label{twoband}
    R_0 = \frac{((n_h\mu_h^2 - n_e\mu_e^2) + (n_h - n_e)(\mu_e\mu_h\mu_0H)^2)}{e((n_e\mu_e + n_h\mu_h)^2 + (n_h - n_e)^2(\mu_e\mu_h\mu_0H)^2)}
\end{equation}
\begin{equation}\label{constraint}
    \rho_{xx}(0) = \frac{1}{e(n_e\mu_e + n_h\mu_h)}
\end{equation}

In Eq. \ref{twoband}, $e$ represents the elementary charge; $n_e$ and $n_h$ are electron and hole carrier concentrations, respectively; and $\mu_e$ and $\mu_h$ are the electron and hole carrier mobilities, respectively. The fit is constrained by Eq. \ref{constraint}, which relates the zero-field resistivity to the concentration and mobilities of the carriers. The difference between the Hall resistivity data and the normal and anomalous Hall contributions is minimized if we assume a dominant extrinsic contribution. Additionally, the magnitudes of the anomalous Hall conductivity and the longitudinal conductivity, as well as the scaling relation between them combined with the anomalous Hall angle, points to a dominant skew-scattering contribution to the anomalous Hall. To confirm reproducibility, transport measurements were performed on additional single crystals from different growth batches. The data shown in Fig. S10 was obtained from a second  sample and shows the same key features as those presented in the main text, demonstrating that the observed topological Hall effect is reproducible across independently grown crystals.

 \begin{figure}[H]
    \centering
    \includegraphics[width=1.0\linewidth]{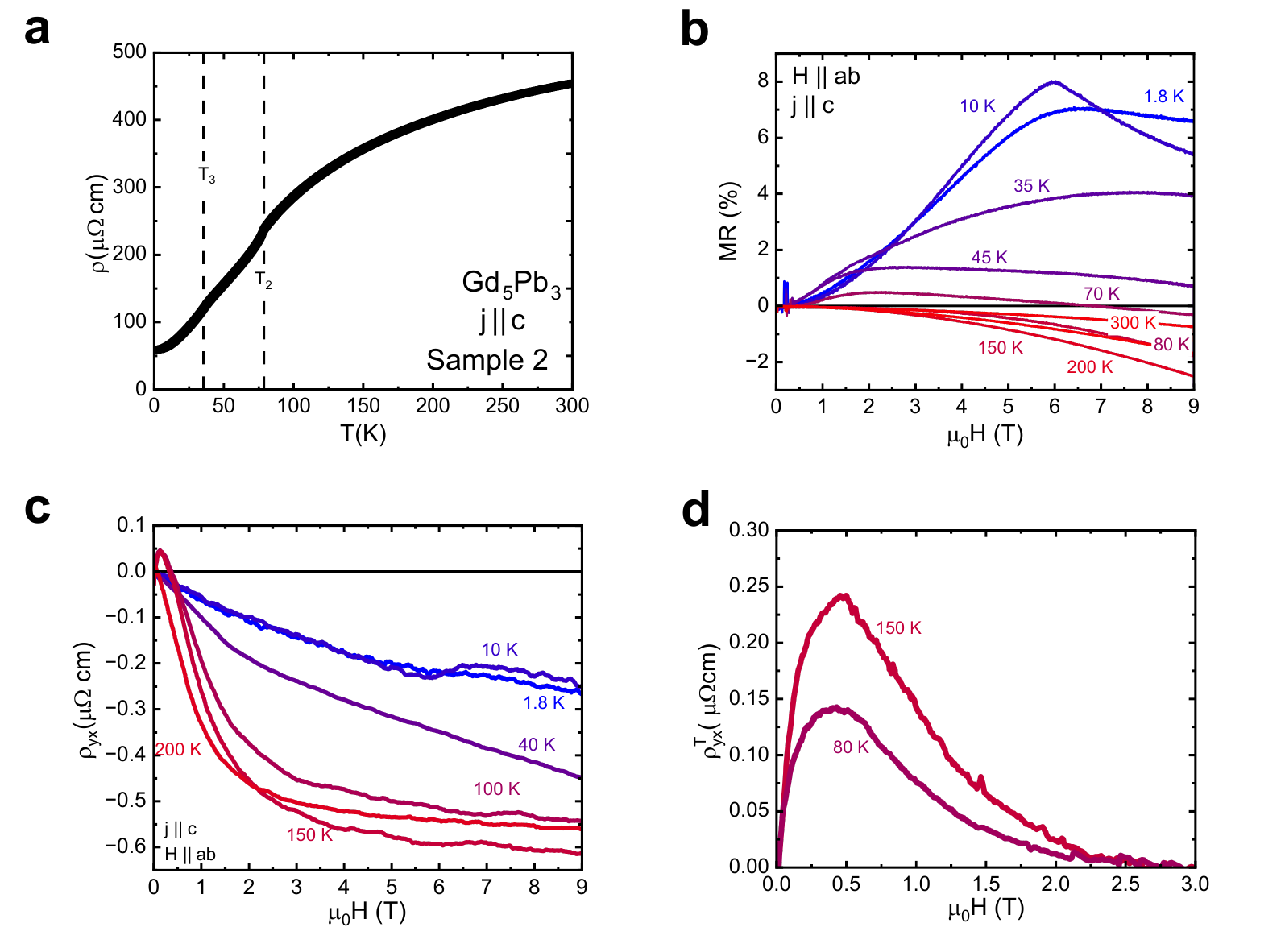}
    \caption{(a) Electrical resistivity as a function of temperature. (b) Magnetoresistance (MR) of Gd5Pb3 as a function of field at different temperatures. (c) Hall resistivity measurements with current j||$c$ and field H||$ab$ for temperatures from 1.8 K to 200 K. (d) The topological Hall resistivity in Gd$_5$Pb$_3$.}
    \label{fitting values second sample}
\end{figure}

\section{Phase Diagram of \texorpdfstring{$\mathbf{Gd_5}\mathbf{Pb_3}$}{Gd5Pb3}}

 \begin{figure}[H]
    \centering
    \includegraphics[width=0.84\linewidth]{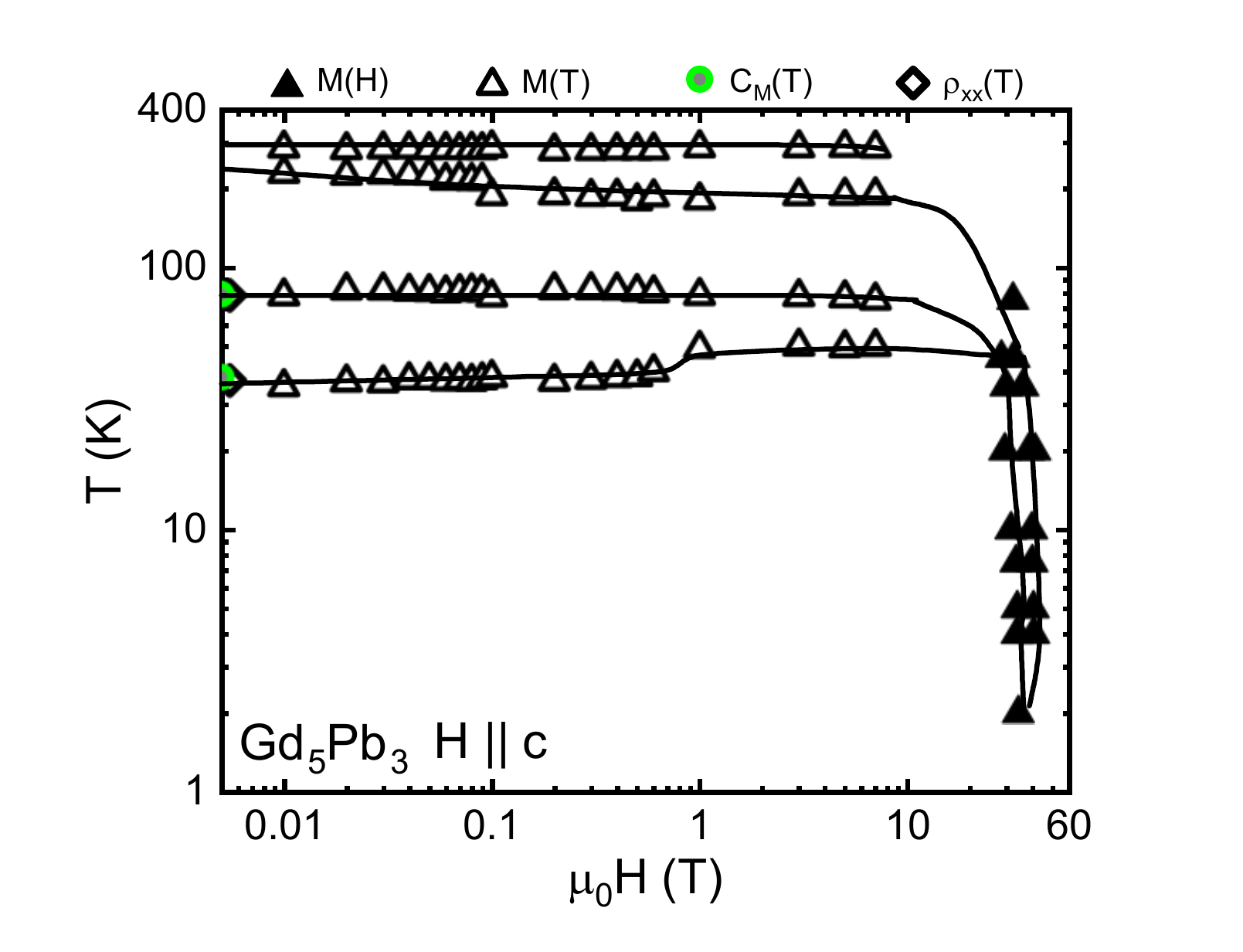}
    \caption{Magnetic $H - T$ phase diagram constructed by peaks from the derivatives of magnetic susceptibility (open triangles), magnetization (filled triangles), magnetic specific heat (green filled circle), resistivity as a function of temperature (open diamonds), resistivity as a function of field (filled diamonds) for $\text{H} || ab$ on a log-log scale. The magnetic specific heat and temperature-dependent resistivity measurements were performed at zero field and have been offset by a small amount to display properly on the log-scale.}
    \label{phasediagram}
\end{figure}
\endgroup
\end{document}